\documentclass[pra,showkeys,showpacs,superscriptaddress,nofootinbib,twocolumn,10pt]{revtex4-1}
\usepackage{graphicx}  
\usepackage{amsthm}
\usepackage{amsmath,amssymb,amsfonts,bbm,MnSymbol,mathrsfs,xfrac}   
\usepackage[breaklinks=true,colorlinks=true,linkcolor=blue,urlcolor=blue,citecolor=blue]{hyperref}
\usepackage{comment}
\usepackage{color}
\usepackage[makeroom]{cancel}

\renewcommand{\[}{\begin{equation}}
\renewcommand{\]}{\end{equation}}
\newcommand{\ket}[1]{|#1\rangle}
\newcommand{\bra}[1]{\langle#1|}

\newcommand{\pro}[2]{|#1\rangle\langle#2|}

\newcommand{\abs}[1]{|#1|}
\newcommand{\ov}[1]{\overline{#1}}
\newcommand{\tr}{\mathrm{tr}}

\newcommand{\norm}[1]{\left\lvert#1\right\rvert}

\newcommand{\A}{\alpha}
\newcommand{\B}{\beta}

\newcommand{\R}{\hat{\rho}}
\newcommand{\de}{d\epsilon}
\newcommand{\bd}{\boldsymbol{d}}
\newcommand{\bA}{{\boldsymbol{\hat{A}}}}
\newcommand{\dbA}{{\Delta\boldsymbol{\hat{A}}}}
\newcommand{\cA}{{\mathscr{A}}}
\newcommand{\bb}{\boldsymbol{b}}
\newcommand{\ba}{\boldsymbol{\alpha}}
\newcommand{\bg}{\boldsymbol{\gamma}}
\newcommand{\be}{{\boldsymbol{\epsilon}}}
\newcommand{\bde}{{\boldsymbol{\mathrm{d}\epsilon}}}
\newcommand{\bdeps}{{\mathrm{d}\!\!\;\boldsymbol{\epsilon}}}
\newcommand{\vectorization}[1]{\mathrm{vec}{[#1]}}
\renewcommand{\Re}{\mathrm{Re}}
\renewcommand{\Im}{\mathrm{Im}}

\newtheorem{definition}{Theorem}
\newtheorem{theorem}[definition]{Theorem}

\begin{document}

\title{
Estimation of Gaussian quantum states}

\author{Dominik \v{S}afr\'{a}nek}
\email{dsafrane@ucsc.edu}
\affiliation{SCIPP and Department of Physics, University of California, Santa Cruz, CA 95064, USA}

\date{\today}

\begin{abstract}
We derive several expressions for the quantum Fisher information matrix (QFIM) for the multi-parameter estimation of multi-mode Gaussian quantum states, the corresponding symmetric logarithmic derivatives, and conditions for saturability of the quantum Cram\'{e}r-Rao bound. This bound determines the ultimate precision with which parameters encoded into quantum states can be estimated. We include expressions for mixed states, for the case when the Williamson decomposition of the covariance matrix is known, expressions in terms of infinite series, and expressions for pure states. We also discuss problematic behavior when some modes are pure, and present a method that allows the use of expressions that are defined only for mixed states, to compute QFIM for states with any number of pure modes.
\end{abstract}

\maketitle

With growing need for precise and cost-effective detectors, quantum metrology paves the way in which the new-era quantum sensors should be designed. This theory gives the ultimate bound on the precision with what parameters of a physical system can be measured by a quantum probe. This bound, called the quantum Cram\'er-Rao bound~\cite{Paris2009a}, might not be generally achievable, either because of practical limitations, or not even in principle~\cite{szczykulska2016multi}. But it still gives an estimate whether a certain design of a new quantum detector is feasible, and whether it can yield a better result than the current technology~\cite{Giovannetti2004a,Giovannetti2006a,zwierz2010general,Giovannetti2011a,Demkowicz2012a}.

To determine this bound, it is necessary to compute the quantum Fisher information matrix~(QFIM), denoted $H(\be)$. Introduced by Holevo~\cite{Holevo2011a}, Helstrom~\cite{helstrom1967minimum,helstrom1976quantum}, by Bures~\cite{Bures1969a}, and later popularized by Braunstein and Caves~\cite{BraunsteinCaves1994a}, QFIM describe limits in distinguishability of infinitesimally close quantum states $\R_{\be}$ and $\R_{\be+\bdeps}$ that differ only by a small variation in parameters that parametrize them. Generally, larger elements of QFIM predict better distinguishability, which therefore leads to a better precision in estimating vector of parameters $\be$.

There has been a wide range of applicability of this quantity, such as optical interferometry used in the detection of gravitational waves~\cite{abbott2016observation} and litography~\cite{boto2000quantum}, applications in thermometry~\cite{Monras2010a,correa2015individual,hofer2017quantum,spedalieri2018thermal}, phase estimation~\cite{Ballester2004a,Monras2006a,Aspachs2008a,demkowicz2009quantum,chin2012quantum,humphreys2013quantum,Nusran2014a,Sparaciari2015a,pezze2017optimal}, estimation of space-time parameters~\cite{nation2009analogue,weinfurtner2011measurement,aspachs2010optimal,kish2017quantum,fink2017experimental}, magnetic fields~\cite{Wasilewski2010a,cai2013chemical,zhang2014fitting,nair2016far}, squeezing parameters~\cite{Milburn1994a,chiribella2006optimal,Gaiba2009a,benatti2011entanglement,safranek2016optimal}, time~\cite{zhang2013criterion,komar2014quantum}, and frequency~\cite{frowis2014optimal,boss2017quantum}.

QFIM has been also used in the description of criticality and quantum phase transitions under the name of `fidelity susceptibility' where they help to describe a sudden change of a quantum state when an external parameter such as magnetic field is varied~\cite{paraoanu1998bures,zanardi2007bures,venuti2007quantum,gu2010fidelity,banchi2014quantum,wu2016geometric,marzolino2017fisher}. Moreover, QFIM measures speed limits on the evolution of quantum states~\cite{margolus1998maximum,taddei2013quantum,del2013quantum,pires2016generalized,deffner2017quantum}, speed limits of quantum computation~\cite{lloyd2000ultimate} and speed limits in charging of batteries~\cite{binder2015quantacell}, it quantifies coherence and entanglement~\cite{hauke2016measuring,girolami2017information,liu2017quantum}, and it provides bounds on irreversibility in open quantum systems~\cite{mancino2018geometrical}.

Many recent experimental setups use Gaussian quantum states. This is because these states are easily created, and easily manipulated. They have been used for example in the aforementioned detection of Gravitational waves~\cite{aasi2013enhanced}, and in Bose-Einstein condensates~\cite{dunningham1998quantum,gross2011atomic,wade2016manipulation}.

Naturally, one often wants to calculate QFIM for Gaussian states. That can be a daunting task, because they appear very complicated in the density matrix formalilsm. Consequently, there has been a lot of effort to make these calculations easier by utilizing the phase-space formalism, which allows to elegantly describe any Gaussian state just by its first and the second moment. Numerous expressions have been derived. For a single parameter estimation, it is QFI for a pure state~\cite{Pinel2012a}, for a nearly pure state~\cite{Friis2015a}, for a single-mode state~\cite{Pinel2013b}, for a two-mode state~\cite{Safranek2015b}, for a multi-mode state~\cite{Monras2013a}, for a multi-mode state in terms of Williamson's decomposition of the covariance matrix~\cite{Safranek2015b}, and for a multi-mode state in terms of infinite series (limit formula)~\cite{Safranek2015b}; for the multi-parameter estimation, it is QFIM for a single-mode state~\cite{Pinel2013b}, for special cases of two-mode states~\cite{marian2016quantum}, for a multi-mode state in terms of tensors~\cite{Gao2014a}, for a multi-mode state in terms of inverses of super-operators~\cite{banchi2015quantum}, and QFIM for a multi-mode state and condition on saturability on the quantum Cram\'er-Rao bound in terms of Williamson's decomposition of the covariance matrix~\cite{nichols2018multiparameter} that differs in form from the single-parameter result of~\cite{Safranek2015b}, and from results published here.

In this paper, we complete the story by deriving several missing expressions for the QFIM for the multi-parameter estimation of multi-mode Gaussian states, expressions for symmetric logarithmic derivatives (SLDs) that determine the optimal measurement, and expressions that determine if the quantum Cram\'er-Rao bound can be saturated. We also discuss and resolve problematic behavior when some of the modes are pure. We devise a regularization procedure, which allows us to use expressions for QFIM that are valid only for mixed Gaussian states to calculate the QFIM for Gaussian states with any number of pure modes.

This paper is structured as follows. We show expressions for QFIM, SLD, and expression determining saturability of the Cram\'er-Rao bound subsequently for cases when 1) all the modes of the Gaussian quantum state are mixed, 2) Williamson's decomposition of the covariance matrix is known, 3) the state is mixed and numerical solutions are enough (limit formulas), and 4) the Gaussian state is pure. In between, at appropriate places, we discuss critical behavior of expressions when some of the modes are pure, associated discontinuous behavior, and we explain the aforementioned regularization procedure. Finally, we illustrate the use of derived formulas on several examples.

\section{Notation and preliminaries}

Lower indices will denote different matrices, while upper indices will denote elements of a matrix. \emph{Bar} as in $\ov{\sigma}$ will denote the complex conjugate, upper index $T$ as in $\sigma^T$ will denote transpose, and $\dag$ as in $\sigma^\dag$ will denote conjugate transpose. $\partial_i\equiv\partial_{\epsilon_i}$ denotes partial derivative with respect to $i$'th element of the vector of estimated parameters $\be=(\epsilon_1,\epsilon_2,\dots)$, $\otimes$ denotes the Kronecker product, $[\cdot,\cdot]$ denotes the commutator, $\{\cdot,\cdot\}$ denotes the anti-commutator, $\tr[\cdot]$ denotes trace of a matrix, and $\vectorization{\cdot}$ denotes vectorization of a matrix, which is defined as a column vector constructed from columns of a matrix as
\[\label{eq:vectorizationdef}
A=\begin{pmatrix}
    a & b \\
    c & d \\
  \end{pmatrix}, \quad
  \vectorization{A}=\begin{pmatrix}
            a \\
            c \\
            b \\
            d \\
          \end{pmatrix}.
\]

We consider a Bosonic system with a set annihilation and creation operators $\{\hat{a}_n,\hat{a}_n^\dag\}$. We collect them into a vector of operators $\bA:=(\hat{a}_1,\dots,\hat{a}_N,\hat{a}_1^\dag,\dots,\hat{a}_N^\dag)^T$, where $N$ denotes the number of modes. Now we can write commutation relations in an elegant form, $[\bA^{m},\bA^{n\dag}]=K^{mn}\mathrm{id}$, where $\mathrm{id}$ denotes the identity operator and
\[\label{def:K}
K=
\begin{bmatrix}
I & 0 \\
0 & -I
\end{bmatrix}
\]
is a constant matrix called \emph{the symplectic form}, and $I$ denotes the identity matrix.

In a Bosonic system\footnote{Although Gaussian states are also defined for Fermionic systems, see e.g. Ref.~\cite{carollo2018symmetric}.} one can define a special class of continuous variable states called \emph{Gaussian states}~\cite{Weedbrook2012a}, $\hat{\rho}$, which are fully characterized by its first moments $\bd^m=\mathrm{tr}\big[\R\bA^m\big]$ (\emph{the displacement vector}) and the second moments $\sigma^{mn}=\mathrm{tr}\big[\hat{\rho}\,\{\dbA^m,\dbA^{n\dag}\}\big]$ (\emph{the covariance matrix}), where $\dbA:=\bA-\bd$. In this form, $\sigma^\dag=\sigma$, and the moments have the following structure,
\[\label{def:first_and_second_moments}
\bd=
\begin{bmatrix}
\boldsymbol{\bg} \\ \overline{\boldsymbol{\bg}}
\end{bmatrix},\quad
\sigma\,=\,\begin{bmatrix}
X & Y \\
\overline{Y} & \overline{X}
\end{bmatrix}.
\]
 These definitions are known as the complex form, and we will use this convention throughout this paper. This description is equivalent to the real form used by some authors~\cite{braunstein2005quantum,Weedbrook2012a,Adesso2014a}. We show how to switch between these descriptions in Appendix~\ref{app:mixed_state}. Both the real and the complex form phase-space represenations of common Gaussian unitaries and Gaussian states are shown in Appendix~\ref{sec:The_phase_space_Gaussian_unitaries}. For more information on the real and the complex form see for example~\cite{Arvind1995a,Safranek2015b,safranek2016optimal,vsafranek2016gaussian}.

 The QFIM is defined as~\cite{Paris2009a}
\[\label{eq:qfi}
H^{ij}\equiv\tfrac{1}{2}\tr[\R\{\hat{\mathscr{L}}_i,\hat{\mathscr{L}}_j\}],
\]
where \emph{symmetric logarithmic derivatives} (SLDs) $\hat{\mathscr{L}}_i$ are defined as operator solutions to equations
\[\label{eq:sld}
\tfrac{1}{2}\{\R,\hat{\mathscr{L}}_i\}=\partial_i\R.
\]
The quantum Cram\'er-Rao bound gives a lower bound on the covariance matrix of estimators,
\[
\mathrm{Cov}(\hat{\be})\geq H^{-1},
\]
meaning that matrix $\mathrm{Cov}(\hat{\be})- H^{-1}$ is a positive semi-definite matrix. This bound is can be saturated when~\cite{szczykulska2016multi,ragy2016compatibility}
\[\label{eq:condition_for_CRbound}
\tr[\R[\hat{\mathscr{L}}_i,\hat{\mathscr{L}}_j]]=0,
\]
and the optimal measurement basis is given by the eigenvectors of the SLDs.

\section{Results}
\textit{\textbf{Mixed states}}

In tasks where we know the covariance matrix and displacement vector of a Gaussian state, it is possible to use results derived in~\cite{Gao2014a} to calculate the QFIM and the symmetric logarithmic derivatives. These results can be expressed in an elegant matrix form: for a Gaussian state $(\bd,\sigma)$, the QFIM and symmetric logarithmic derivatives can be calculated as
\begin{align}
H^{ij}(\be)&=\frac{1}{2}\vectorization{\partial_i\sigma}^\dag\mathfrak{M}^{-1}\vectorization{\partial_j\sigma}+2\partial_i\bd^\dag\sigma^{-1}\partial_j\bd,\label{eq:mixed_QFI}\\
\hat{\mathscr{L}}_i(\be)&=\dbA^\dag\cA_i\dbA-\frac{1}{2}\tr[\sigma\cA_i]+2\dbA^\dag\sigma^{-1}\partial_i{\boldsymbol{d}},\label{eq:SLD_mixed}
\end{align}
where
\[
\mathfrak{M}=\ov{\sigma}\otimes\sigma-K\otimes K,\quad
\vectorization{\cA_i}=\mathfrak{M}^{-1}\vectorization{\partial_i\sigma}.
\]
These results represent the Gaussian version of the expressions published in Ref.~\cite{vsafranek2018simple}.

The above formulas require differentiating the displacement vector and the covariance matrix, and inverting two matrices. However, they cannot be used when at least one of the modes is in a pure state (at least not without any modification), because $\mathfrak{M}$ is not invertible in that case. We will discuss this issue later, and show how this can be resolved (see Eq.~\eqref{eq:any_state_QFIM}).

Proof of how results of~\cite{Gao2014a} transform into this elegant matrix form can be found in Appendix~\ref{app:mixed_state}.

Further, we derive expression that determines saturability of the quantum Cram\'er-Rao bound,
\[\label{eq:condition_QCR}
\begin{split}
\tr[\R[\hat{\mathscr{L}}_i,\hat{\mathscr{L}}_j]]&\!=\!\vectorization{\partial_i\sigma}^\dag\mathfrak{M}^{-1}(\ov{\sigma}\!\otimes\! K\!-\!K\!\otimes\!\sigma)\mathfrak{M}^{-1}\vectorization{\partial_j\sigma}\\
&+4\partial_i\bd^\dag\sigma^{-1}K\sigma^{-1}\partial_j\bd.
\end{split}
\]
The derivation can be found in Appendix~\ref{app:saturability}.

\textit{\textbf{When the Williamson's decomposition of the covariance matrix is known}}

There are situations when we know the Williamson's decomposition of the covariance matrix. This is for example when we have control over preparation of the initial state, and we are planning to estimate parameters encoded into this state by a Gaussian unitary channel. This is a case when it is useful to simplify our calculations of QFIM by making use of this decomposition, without the actual need of calculating this decomposition. In other cases, it might be still numerically more efficient to calculate this decomposition, instead of inverting the large matrix $\mathfrak{M}$ needed for Eq.~\eqref{eq:mixed_QFI}.

According to the Williamson's theorem~\cite{Williamson1936a,deGosson2006a,simon1999congruences} any positive-definite matrix can be diagonalized by symplectic matrices, $\sigma=SDS^\dag$. We show how to do that explicitly in Appendix~\ref{app:diagonalization}. $D=\mathrm{diag}(\lambda_1,\dots,\lambda_N,\lambda_1,\dots,\lambda_N)$ is a diagonal matrix consisting of \emph{symplectic eigenvalues}, which are defined as the positive eigenvalues of the matrix $K\sigma$. It follows from Heisenberg uncertainty relations that for all $k$, $\lambda_k\geq 1$. $S$ is a \emph{symplectic matrix} satisfying the defining property of the complex form of the real symplectic group $Sp(2N,\mathbb{R})$,
\[\label{def:structure_of_S}
S=
\begin{bmatrix}
\A & \B \\
\ov{\B} & \ov{\A}
\end{bmatrix},\ \ SKS^\dag=K.
\]

We define matrices $P_i:=S^{-1}\partial_i{S}$, which are elements of the Lie algebra associated with the symplectic group, satisfying the defining properties of this algebra,
\[\label{def:P_1}
P_i=
\begin{bmatrix}
R_i & Q_i \\
\ov{Q}_i & \ov{R}_i
\end{bmatrix},\ \ P_iK+KP_i^\dag=0.
\]

Common symplectic matrices in the complex form, representing for example a squeezing operation, phase-change, or a beam-splitter, can be found in Appendix~\ref{sec:The_phase_space_Gaussian_unitaries} or in more detail in Ref.~\cite{safranek2016optimal} and Section II of Ref.~\cite{vsafranek2016gaussian}.

Rewriting Eq.~\eqref{eq:mixed_QFI} in terms of the Williamson's decomposition of the covariance matrix, switching to element-wise notation, and simplifying using identities~\eqref{def:structure_of_S} and~\eqref{def:P_1}, we derive we derive an analytical expression for the quantum Fisher information matrix of Gaussian states in terms of the Williamson's decomposition of the covariance matrix,
\begin{align}
H^{ij}(\be)&\!\!=\!\!\!\!\sum_{k,l=1}^{N}\!\!\frac{(\lambda_k\!-\!\lambda_l)^2}{\lambda_k\lambda_l\!-\!1}\Re[\ov{R_{i}}^{kl}\!R_{j}^{kl}]+\frac{(\lambda_k\!+\!\lambda_l)^2}{\lambda_k\lambda_l\!+\!1}\Re[\ov{Q_{i}}^{kl}\!Q_{j}^{kl}]\nonumber\\
&\!+\sum_{k=1}^{N}\frac{\partial_i\lambda_k\partial_j\lambda_k}{\lambda_k^2-1}+2\partial_i\boldsymbol{d}^\dag\sigma^{-1}\partial_j\boldsymbol{d}.\label{eq:multimode_QFI}
\end{align}
$\Re$ denotes the real part, $R_i=\A^\dag\partial_i{\A}-\overline{\B^\dag\partial_i{\B}}$ is a skew-Hermitian and $Q_i=\A^\dag\partial_i{\B}-\overline{\B^\dag\partial_i{\A}}$ a (complex) symmetric matrix. We note that $\sigma^{-1}=KSD^{-1}S^\dag K$, which follows from Eq.~\eqref{def:structure_of_S} and properties of $K$. The above formula represents a multi-parameter generalization of the result for a single-parameter estimation published in Ref.~\cite{Safranek2015b}. The full derivation can be found in Appendix~\ref{app:williamson_QFI}.

The above formula can be used even for states with some pure modes, defined by $\lambda_k=1$ for some mode $k$. We will show how soon.

When none of the modes are pure, i.e., all symplectic eigenvalues are larger than one, we can rewrite Eq.~\eqref{eq:multimode_QFI} in a very elegant way. Defining Hermitian matrix $\widetilde{R}_i^{kl}:=\frac{\lambda_k-\lambda_l}{\sqrt{\lambda_k\lambda_l-1}}R_i^{kl}$, symmetric matrix $\widetilde{Q}_i^{kl}:=\frac{\lambda_k+\lambda_l}{\sqrt{\lambda_k\lambda_l+1}}Q_i^{kl}$, and diagonal matrix $L:=\mathrm{diag}(\lambda_1,\dots,\lambda_N)$, QFIM can be written as
\[\label{eq:exact_multimode_compact}
\begin{split}
H^{ij}(\be)&=\frac{1}{2}\tr\big[\widetilde{R}_i\widetilde{R}_j^\dag+\widetilde{R}_j\widetilde{R}_i^\dag+\widetilde{Q}_i\widetilde{Q}_j^\dag+\widetilde{Q}_j\widetilde{Q}_i^\dag\big]\\
&+\tr\big[(L^2-I)^{-1}\partial_iL\partial_jL\big]+2\partial_i\boldsymbol{d}^\dag\sigma^{-1}\partial_j\boldsymbol{d}.
\end{split}
\]
Looking at the derivatives in each term, we can conclude that QFIM consists of three qualitatively different terms: the first term is connected to the change of orientation and squeezing of the Gaussian state with small variations in $\be$, the second to the change of purity, and the third to the change of displacement.

Similarly, we derive expression for the symmetric logarithmic derivative,
\[\label{eq:SLD_mixed_williamson}
\hat{\mathscr{L}}_i(\be)=\dbA^\dag \!(S^{-1})^\dag W_i S^{-1}\!\dbA-\sum_{k=1}^N\frac{\lambda_k\partial_i\lambda_k}{\lambda_k^2-1}+2\dbA^\dag\!\sigma^{-1}\partial_i{\boldsymbol{d}},
\]
where $W_i$ is a Hermitian matrix of form
\[
\begin{split}
W_i&=\begin{bmatrix}
W_{Xi} & W_{Yi} \\
\overline{W_{Yi}} & \overline{W_{Xi}}
\end{bmatrix},\\
W_{Xi}^{kl}
&=-\frac{\lambda_k-\lambda_l}{\lambda_k\lambda_l-1}R_i^{kl}+\frac{\partial_i\lambda_k}{\lambda_k^2-1}\delta^{kl},\\
W_{Yi}^{kl}&=\frac{\lambda_k+\lambda_l}{\lambda_k\lambda_l+1}Q_i^{kl}.
\end{split}
\]
Further, we derive expression that determines saturability of the quantum Cram\'er-Rao bound,
\[\label{eq:multimode_condition}
\begin{split}
&\tr[\R[\hat{\mathscr{L}}_i,\hat{\mathscr{L}}_j]]=4\partial_i\bd^\dag\sigma^{-1}K\sigma^{-1}\partial_j\bd\\
&+\!\!\sum_{k,l=1}^{N}\!\!\frac{2i(\lambda_k\!+\!\lambda_l)^3}{(\lambda_k\lambda_l\!+\!1)^2}\Im[\ov{Q_{i}}^{kl}\!Q_{j}^{kl}]-\frac{2i(\lambda_k\!-\!\lambda_l)^3}{(\lambda_k\lambda_l\!-\!1)^2}\Im[\ov{R_{i}}^{kl}\!R_{j}^{kl}].\\
\end{split}
\]
The derivation can be found in Appendix~\ref{app:williamson_QFI}. Alternative but equivalent forms of the above expressions are also published in Ref.~\cite{nichols2018multiparameter}.

\textit{\textbf{When some of the modes are pure}}

Eqs.~(\ref{eq:mixed_QFI},\ref{eq:SLD_mixed},\ref{eq:condition_QCR},\ref{eq:multimode_QFI},\ref{eq:SLD_mixed_williamson},\ref{eq:multimode_condition}) are not well defined for states that have at least one mode in a pure state, i.e., $\lambda_k=1$ for some mode $k$, which also results in matrix $\mathfrak{M}$ not being invertible. It has been shown~\cite{vsafranek2017discontinuities}, and we explain it in detail in Appendix~\ref{app:pops}, that there are two unique ways of defining QFIM at these problematic points, depending on the quantity we want to obtain.

First, we will illustrate this on Eq.~\eqref{eq:multimode_QFI}. To obtain the QFIM, which is defined~\cite{Paris2009a} through the symmetric logarithmic derivatives $\mathscr{L}_i$ as $H^{ij}\equiv\tfrac{1}{2}\tr[\R\{\hat{\mathscr{L}}_i,\hat{\mathscr{L}}_j\}]$, we define problematic terms as
\[\label{def:problematic_points}
\frac{(\lambda_k\!-\!\lambda_l)^2}{\lambda_k\lambda_l\!-\!1}=0,\quad
\frac{\partial_i\lambda_k\partial_j\lambda_k}{\lambda_k^2-1}=0,
\]
for $\be$ such that $\lambda_k(\be)=\lambda_l(\be)=1$.

The continuous quantum Fisher information matrix (cQFIM) is defined~\cite{vsafranek2017discontinuities} as four-times the Bures metric as $H_c^{ij}\equiv 4g^{ij}$, where the Bures metric is defined by $\sum_{i,j}g^{ij}(\be)\mathrm{d}\epsilon_i\mathrm{d}\epsilon_j\equiv 2\big(1-\sqrt{\mathcal{F}(\R_\be,\R_{\be+\bde})}\big)$, and ${\mathcal{F}({\R}_{1},{\R}_{2})=\big(\tr\sqrt{\sqrt{{\R}_{1}}\,{\R}_{2}\,\sqrt{{\R}_{1}}}\big)^{2}}$  denotes the Uhlmann's fidelity~\cite{Uhlmann1976a}.
To obtain the cQFIM, we define
\[\label{def:problematic_points2}
\frac{(\lambda_k\!-\!\lambda_l)^2}{\lambda_k\lambda_l\!-\!1}=0,\quad
\frac{\partial_i\lambda_k\partial_j\lambda_k}{\lambda_k^2-1}=\partial_i\partial_j\lambda_k,
\]
for $\be$ such that $\lambda_k(\be)=\lambda_l(\be)=1$.

QFIM and cQFIM are identical everywhere, apart from those problematic points~\cite{vsafranek2017discontinuities}. At those points, QFIM can be discontinuous, while cQFIM is in some sense continuous~\cite{vsafranek2017discontinuities}. Defining Hessian matrix $\mathcal{H}_k^{ij}:=\partial_i\partial_j\lambda_k$, we can use the above equations to write relation
\[\label{eq:QFIM_cQFIM_relation}
H_c(\be)=H(\be)+\!\!\!\!\!\!\sum_{k:\lambda_k(\be)=1}\!\!\!\!\!\!\mathcal{H}_k(\be).
\]
By writing $k:\lambda_k(\be)=1$ we mean that the sum goes only over values of $k$ for which $\lambda_k(\be)=1$. For any $k$ such that $\lambda_k(\epsilon)=1$, $\mathcal{H}_k(\be)$ is positive semi-definite, and we can therefore write $H_c\geq H$. $H_c= H$ if and only if for all $k$ such that $\lambda_k=1$, $\mathcal{H}_k=0$.

Similarly, in Eqs.~\eqref{eq:SLD_mixed_williamson} and~\eqref{eq:multimode_condition} we define $\frac{\lambda_k\!-\!\lambda_l}{\lambda_k\lambda_l\!-\!1}=\frac{(\lambda_k\!-\!\lambda_l)^3}{(\lambda_k\lambda_l\!-\!1)^2}=\frac{\partial_i\lambda_k}{\lambda_k^2-1}:=0$ for $\be$ such that $\lambda_k(\be)=\lambda_l(\be)=1$.

\textit{\textbf{Regularization procedure}}

Now we will show how to treat cases when some modes are pure in general, not limiting ourselves to already resolved case of Eq.~\eqref{eq:multimode_QFI}. We can devise a regularization procedure that will allow us to use expressions that work only for states where all the modes are mixed, such as Eq.~\eqref{eq:mixed_QFI}, to compute the QFIM for any state. Similar method has been already used for regularizing QFIM for non-Gaussian states~\cite{vsafranek2017discontinuities}.

It goes as follows. First we multiply the covariance matrix by regularization parameter $\nu>1$, and use some expression, such as Eq.~\eqref{eq:mixed_QFI}, to  calculate the QFIM for state $(\bd,\nu\sigma)$. Then we perform limit $\nu\rightarrow 1$. The resulting value will represent the correct QFIM for state $(\bd,\sigma)$.

To prove that, however, we have to check that this limit leads to the proper definition of the problematic points, as given by Eq.~\eqref{def:problematic_points}. We take Eq.~\eqref{eq:multimode_QFI} as a study case, but because this formula is general, the result will be valid for any other expression for QFIM. When covariance matrix $\sigma$ has symplectic eigenvalues $\lambda_k$, covariance matrix $\nu\sigma$ has symplectic eigenvalues $\nu\lambda_k$. Sympletic matrices from the decompositions of $\sigma$ and $\nu\sigma$ are identical. Parameter $\nu$ therefore appears only as a modification of symplectic eigenvalues, which we will take advantage of. Assuming $\lambda_k(\be)=\lambda_l(\be)=1$ and performing the limit, both problematic terms are set to zero by taking the limit, $\lim_{\nu\rightarrow1}\frac{(\nu\lambda_k-\nu\lambda_l)^2}{\nu\lambda_k\nu\lambda_l-1}=\lim_{\nu\rightarrow1}\frac{0}{\nu^2-1}=0$, $\lim_{\nu\rightarrow1}\frac{(\nu\partial_i{\lambda}_k)^2}{(\nu\lambda_k)^2-1}=\lim_{\nu\rightarrow1}\frac{0}{\nu^2-1}=0$, which is exactly the definition, Eq.~\eqref{def:problematic_points}, that we wanted.  $\partial_i{\lambda}_k(\be)=0$, because ${\lambda}_k(\be)$ achieves a local minimum at point $\be$ when $\lambda_k(\be)=1$.

This method therefore leads to the correct value of the QFIM, and we can write expression for the QFIM for any Gaussian quantum state as
\[\label{eq:regularization_procedure}
H(\be)\equiv H(\bd(\be),\sigma(\be))=\lim_{\nu\rightarrow1}H\big(\bd(\be),\nu\sigma(\be)\big).
\]

Applying this result to Eq.~\eqref{eq:mixed_QFI}, QFIM for any state can be computed as
\[\label{eq:any_state_QFIM}
\begin{split}
H^{ij}(\be)&=\lim_{\nu\rightarrow1} \frac{1}{2}\vectorization{\partial_i\sigma}^\dag(\nu^2\ov{\sigma}\otimes\sigma-K\otimes K)^{-1}\vectorization{\partial_j\sigma}\\
&+2\partial_i\bd^\dag\sigma^{-1}\partial_j\bd.
\end{split}
\]

Expression for the cQFIM can obtained by combining Eq.~\eqref{eq:regularization_procedure} with Eq.~\eqref{eq:QFIM_cQFIM_relation}.

Similarly, we have
\[\label{eq:SLD_regularization}
\hat{\mathscr{L}}_i(\be)=\lim_{\nu\rightarrow1}\dbA^\dag\cA_{i\nu}\dbA-\frac{1}{2}\tr[\sigma\cA_{i\nu}]+2\dbA^\dag\sigma^{-1}\partial_i{\boldsymbol{d}},
\]
and
\[\label{eq:condition_QCRregularization_procedure2}
\begin{split}
&\tr[\R[\hat{\mathscr{L}}_i,\hat{\mathscr{L}}_j]]=4\partial_i\bd^\dag\sigma^{-1}K\sigma^{-1}\partial_j\bd\\
&+\lim_{\nu\rightarrow1}\vectorization{\partial_i\sigma}^\dag\mathfrak{M}_{\nu}^{-1}(\ov{\sigma}\!\otimes\! K\!-\!K\!\otimes\!\sigma)\mathfrak{M}_{\nu}^{-1}\vectorization{\partial_j\sigma},
\end{split}
\]
where $\vectorization{\cA_{i\nu}}=\mathfrak{M}_{\nu}^{-1}\vectorization{\partial_i\sigma}$ and $\mathfrak{M}_{\nu}=\nu^2\ov{\sigma}\otimes\sigma-K\otimes K$.

\textit{\textbf{Limit formula}}

We presented exact analytical expressions for QFIM, however, in some cases a numerical value that approximates the exact value to any desired precision is enough. Defining matrix $A:=K\sigma$, and generalizing procedure derived in Ref.~\cite{Safranek2015b} for a single-parameter estimation, we derive the limit expression for the QFIM,
\[\label{eq:limit_formula}
H^{ij}(\be)=\frac{1}{2}\sum_{n=1}^\infty\tr\big[A^{-n}\partial_i{A}A^{-n}\partial_j{A}]+2\partial_i\boldsymbol{d}^\dag\sigma^{-1}\partial_j\boldsymbol{d}.
\]
This expression is proved in the next section, by showing its relation to Eq.~\eqref{eq:mixed_QFI}.

Note that although the above infinite series converges even when some symplectic eigenvalue are equal to one, for such cases it is not absolutely convergent and it does not give the correct expression for the QFIM. This can be shown by careful analysis of the elements in the series given by Eq.~\eqref{eq:expression_in_terms_of_P}, and it was explained in detail in Ref.~\cite{Safranek2015b}. The correct expression for cases when some of the modes are pure can be obtained by combining the above equation with the regularization procedure, Eq.~\eqref{eq:regularization_procedure}.

In applications, we would like to take just a few elements of the series, and believe that their sum well approximates the QFIM. To estimate the error when doing this, we define remainder of the series as $R_M^{ij}:=\frac{1}{2}\sum_{n=M+1}^\infty\tr\big[A^{-n}\!\partial_i{A}A^{-n}\!\partial_j{A}]$. As shown in Appendix~\ref{app:Remainder}, this remainder is bounded,
\[\label{eq:remainder}
|R_M^{ij}|\leq\frac{\sqrt{\mathrm{tr}[(A\partial_iA)^2]}\sqrt{\mathrm{tr}[(A\partial_jA)^2]}}{2\lambda_{\mathrm{min}}^{2(M+1)}(\lambda_{\mathrm{min}}^2-1)},
\]
where $\lambda_{\mathrm{min}}:=\min_{k}\{\lambda_k\}$ is the smallest symplectic eigenvalue of the covariance matrix $\sigma$. The right hand side therefore represents the maximal error when calculating the QFIM by using the first $M$ elements of the series, Eq.~\eqref{eq:limit_formula}.

We can derive similar limit expressions for the SLD and for $\tr[\R[\hat{\mathscr{L}}_i,\hat{\mathscr{L}}_j]]$ by using expression
\[
\cA_i=\sum_{n=1}^\infty A^{-n}\partial_iA A^{-n}K,
\]
derived towards the end of Appendix~\ref{app:mixed_state}.

\textit{\textbf{Relations between different expressions for QFIM}}

Here we show how the limit expression, Eq.~\eqref{eq:limit_formula}, relates to Eqs.~\eqref{eq:mixed_QFI} and~\eqref{eq:multimode_QFI}. Relation between Eqs.~\eqref{eq:mixed_QFI} and~\eqref{eq:multimode_QFI} is shown in Appendix~\ref{app:williamson_QFI}.

To obtain Eq.~\eqref{eq:mixed_QFI}, we use $\sigma^\dag=\sigma$, properties of vectorization, $\tr[A^\dag B]=\vectorization{A}^\dag\vectorization{B}$, and properties of Kronecker product, $(AB)\otimes(A'B)=(A\otimes A')(B\otimes B')$, $(C^T\otimes A)\vectorization{B}=\vectorization{ABC}$, to transform the infinite sum in Eq.~\eqref{eq:limit_formula} into a Neumann series that can be evaluated,
\[
\begin{split}
&\sum_{n=1}^\infty\tr\big[A^{-n}\partial_i{A}A^{-n}\partial_j{A}]\\
&=\vectorization{\partial_i\sigma}^\dag\bigg(\sum_{n=0}^\infty(\ov{A}\otimes A)^{-n}\bigg)\big(\ov{\sigma}\otimes\sigma\big)^{-1}\vectorization{\partial_j\sigma}\\
&=\vectorization{\partial_i\sigma}^\dag(I-\ov{A}^{-1}\otimes A^{-1})^{-1}\big(\ov{\sigma}\otimes\sigma\big)^{-1}\vectorization{\partial_j\sigma}\\
&=\vectorization{\partial_i\sigma}^\dag(\ov{\sigma}\otimes\sigma-K\otimes K)^{-1}\vectorization{\partial_j\sigma},
\end{split}
\]
which gives Eq.~\eqref{eq:mixed_QFI}.

Using identity
\[\label{eq:expression_in_terms_of_P}
\begin{split}
&\tr[A^{-n}\partial_i{A}A^{-n}\partial_j{A}]=2\tr[D^{-n+1}\!K^{-n+1}\!P_{i}D^{-n+1}\!K^{-n+1}\!P_{j}]\\
&-\tr[D^{-n+2}\!K^n\!P_{i}D^{-n}\!K^n\!P_{j})]
-\tr[D^{-n+2}\!K^n\!P_{j}D^{-n}\!K^n\!P_{i})]\\
&+\tr[D^{-n}\partial_i{D}D^{-n}\partial_j{D}]
\end{split}
\]
and changing to element-wise notation, the infinite sum~\eqref{eq:limit_formula} turns out to be geometric series in powers of $\lambda_k$'s, which can be evaluated. Then, using $R_i^{kl}=-\ov{R}_i^{lk}$, $Q_i^{kl}=Q_i^{lk}$ which follows from Eq.~\eqref{def:P_1}, we prove that Eq.~\eqref{eq:limit_formula} simplifies to Eq.~\eqref{eq:multimode_QFI}.

\textit{\textbf{Pure states}}

Combining Eq.~\eqref{eq:limit_formula}, the regularization procedure~\eqref{eq:regularization_procedure}, and $A^2(\be)=I$ (which holds for pure states because for them, $\lambda_k(\be)=1$ for all $k$), we obtain the well-known result for pure states~\cite{Pinel2012a},
\[\label{eq:pure_non-elegant}
H^{ij}(\be)=\frac{1}{4}\mathrm{tr}[\sigma^{-1}\partial_i\sigma\sigma^{-1}\partial_j\sigma]+2\partial_i\boldsymbol{d}^\dag\sigma^{-1}\partial_j\boldsymbol{d}.
\]
It is important to stress that although this expression is defined for any state, it can be applied only for states that are pure at point $\be$. If the state becomes mixed when $\be$ is slightly varied, i.e., when $\partial_i\partial_j\lambda_k(\be)\neq 0$ for some $k$ (see Appendix~\ref{app:pops}, and Ref.~\cite{vsafranek2017discontinuities}), QFIM at this varied parameter $H(\be+\bdeps)$ has to be calculated using some other formula (for example, Eqs.~\eqref{eq:mixed_QFI}, \eqref{eq:multimode_QFI}, or \eqref{eq:any_state_QFIM}), and one finds that in that case, function $H^{ij}$ is discontinuous at point $\be$.

To obtain cQFIM for states that are pure at point $\be$, i.e., $A^2(\be)=I$, we can use expression
\[\label{eq:pure_elegant}
\begin{split}
H_c^{ij}(\be)&=\frac{1}{4}\big(2\mathrm{tr}[\sigma^{-1}\partial_i\partial_j\sigma]-\mathrm{tr}[\sigma^{-1}\partial_i\sigma\sigma^{-1}\partial_j\sigma]\big)\\
&+2\partial_i\boldsymbol{d}^\dag\sigma^{-1}\partial_j\boldsymbol{d},
\end{split}
\]
To prove this expression, one needs to utilize the Williamson's decomposition of the covariance matrix, find~$\mathrm{tr}[\sigma^{-1}\partial_i\partial_j\sigma]$ and $\mathrm{tr}[\sigma^{-1}\partial_i\sigma\sigma^{-1}\partial_j\sigma]$ in terms of matrices $K$, $P_i=S^{-1}\partial_i S,$ and $P_{ij}:=S^{-1}\partial_i\partial_jS$ (which will give expressions similar to Eq.~\eqref{eq:expression_in_terms_of_P}), and use Eqs.~\eqref{def:structure_of_S},~\eqref{def:P_1}, and $P_{ij}K+P_iKP_j^\dag+P_jKP_i^\dag+KP_{ji}^\dag=0$. When applied to both Eq.~\eqref{eq:pure_non-elegant} and Eq.~\eqref{eq:pure_elegant}, we find that Eq. ~\eqref{eq:pure_elegant} gives the same expression as Eq.~\eqref{eq:pure_non-elegant}, plus an additional factor given by the second part of Eq.~\eqref{eq:QFIM_cQFIM_relation}. This proves that Eq.~\eqref{eq:pure_elegant} represents the cQFIM for pure states.

Note that Eqs.~\eqref{eq:pure_non-elegant} and~\eqref{eq:pure_elegant} can be further simplified by using $\sigma^{-1}=K\sigma K$ and $\partial_i AA=-A\partial_i A$, which follows from $A^2(\be)=I$.

Finally, we derive symmetric logarithmic derivatives for pure states
\[\label{eq:sld_pure}
\hat{\mathscr{L}}_i(\be)=\frac{1}{2}\dbA^\dag\!\sigma^{-1}\! \partial_i\sigma \sigma^{-1}\dbA+2\dbA^\dag\!\sigma^{-1}\partial_i{\boldsymbol{d}},
\]
and
\[\label{eq:pure_state_condition}
\tr[\R[\hat{\mathscr{L}}_i,\hat{\mathscr{L}}_j]]\!=\!\frac{1}{4}\tr[K\!\sigma[K\partial_i\sigma,K\partial_j\sigma]]
+4\partial_i\bd^\dag\!\sigma^{-1}\!K\!\sigma^{-1}\!\partial_j\bd.
\]
The derivation can be found in Appendix~\ref{app:pure_state_condition}.

\section{Examples}
Here we illustrate the derived formulas on several examples. As shown in~\cite{safranek2016optimal}, Gaussian unitary operations, which are generated via an exponential map with the exponent at most quadratic in the field operators~\cite{Weedbrook2012a}, can be parameterized by matrix $W$ and vector $\ba$ as
\[\label{def:Gaussian_unitary}
\hat{U}=\exp\big(\tfrac{i}{2}\bA^\dag W \bA+\bA^\dag K \ba\big).
\]
In case of $W=0$, this operator corresponds to the Weyl displacement operator, while for $\ba=0$ we obtain purely quadratic transformations such as the phase-changing operator, one- and two-mode squeezing operators, or mode-mixing operators, depending on the particular structure of $W$. The first and the second moments of the transformed density matrix $\hat{\rho}'=\hat{U}\hat{\rho}\hat{U}^\dag$ are computed as
\[\label{def:transformation}
\bd'=S\bd+\bb,\ \ \sigma'=S\sigma S^\dag,
\]
where the symplectic matrix and the displacement are given by
\[\label{eq:S_and_b}
S=e^{iKW},\ \
\bb=\Big(\!\int_0^1e^{iKWt}\mathrm{d}t\!\Big)\ \!\ba.
\]
The states in the following examples are generated using the above transformations, usually applied on a thermal state. Their form in the phase-space formalism is explicitly computed in Appendix~\ref{sec:The_phase_space_Gaussian_unitaries}, or in more detail in Ref.~\cite{safranek2016optimal} and in Chapter II~of Ref.~\cite{vsafranek2016gaussian}.

Although every formula demonstrated here can be used to calculate QFIM for Gaussian states of any number of modes, for simplicity we choose only single- and two-mode states. However, we point out that for a single- and two-mode Gaussian states it is often better to use expressions valid for these specific number of modes~\cite{Pinel2013b,Safranek2015b}.

\textit{\textbf{Mixed states}}

Let us consider estimation of a squeezing parameter $r$ and inverse temperature $\beta$ from a squeezed thermal state, $\R=\hat{S}(r)\R_{\mathrm{th}}(\beta)\hat{S}^\dag(r)$, where $\hat{S}(r)$ denotes the squeezing operator and $\R_{\mathrm{th}}(\beta)=\frac{1}{Z}\exp(-\beta\hat{n})$ is the thermal state. $Z$ denotes the partition function and $\hat{n}$ denotes the number operator. The final state can be expressed via the first and the second moment as
\[\label{eq:first_example_moments}
\bd=\begin{pmatrix}
        0 \\
        0 \\
      \end{pmatrix}
,
\quad \sigma=\lambda\begin{pmatrix}
\cosh 2r & -\sinh 2r \\
-\sinh 2r & \cosh 2r
\end{pmatrix},
\]
where $\lambda=\coth\tfrac{\beta}{2}$. We compute
\begin{widetext}
\begin{gather}
\mathfrak{M}^{-1}=\frac{\lambda^2}{2(\lambda^4-1)}\begin{pmatrix}
               \cosh 4r+1+\tfrac{2}{\lambda^2} & \sinh
               4r & \sinh
               4r & \cosh 4r-1 \\
               \sinh
               4r & \cosh 4r+1-\tfrac{2}{\lambda^2} & \cosh 4r-1 & \sinh
               4r \\
               \sinh
               4r & \cosh 4r-1& \cosh 4r+1-\tfrac{2}{\lambda^2} & \sinh
               4r \\
               \cosh 4r-1 & \sinh
               4r & \sinh
               4r & \cosh 4r+1+\tfrac{2}{\lambda^2} \\
             \end{pmatrix},\\
\vectorization{\partial_\beta\sigma}=\frac{\lambda^2-1}{2}\begin{pmatrix}
            -\cosh
               2r \\
            \sinh
               2r \\
            \sinh
               2r \\
            -\cosh
               2r \\
          \end{pmatrix}, \quad
\vectorization{\partial_r\sigma}=2\lambda\begin{pmatrix}
            \sinh
               2r \\
            -\cosh
               2r \\
            -\cosh
               2r \\
            \sinh
               2r \\
          \end{pmatrix}.\nonumber
\end{gather}
QFIM is calculated from Eq.~\eqref{eq:mixed_QFI},
\[\label{eq:example1exact}
H(\beta,r)=\frac{1}{2}\begin{pmatrix}
\vectorization{\partial_\beta\R}^\dag\mathfrak{M}^{-1}\vectorization{\partial_\beta\R} &\vectorization{\partial_\beta\R}^\dag\mathfrak{M}^{-1}\vectorization{\partial_r\R} \\
 \vectorization{\partial_r\R}^\dag\mathfrak{M}^{-1}\vectorization{\partial_\beta\R} & \vectorization{\partial_r\R}^\dag\mathfrak{M}^{-1}\vectorization{\partial_r\R} \\
\end{pmatrix}=
\begin{pmatrix}
         \frac{\lambda^2-1}{4} & 0 \\
         0 & \frac{4\lambda^2}{\lambda^2+1} \\
       \end{pmatrix}.
\]
\end{widetext}

From Eq.~\eqref{eq:condition_QCR} we derive $\tr[\R[\hat{\mathscr{L}}_i,\hat{\mathscr{L}}_j]]=0$ for all $i,j=\beta,r$, which according to Eq.~\eqref{eq:condition_for_CRbound} means that the quantum Cram\'er-Rao bound is achievable.

\textit{\textbf{When the Williamson's decomposition of the covariance matrix is known}}

Let the initial state be a coherent state $\ket{\alpha}$, which is given by
\[
\bd_0=\begin{pmatrix}
        \alpha \\
        \ov{\alpha} \\
      \end{pmatrix}
,
\quad \sigma_0=\begin{pmatrix}
1 & 0 \\
0 & 1
\end{pmatrix},
\]
and we use this state to probe squeezing channel with subsequent phase-change, which transforms the state to $\R=\hat{R}(\theta)\hat{S}(r)\pro{\alpha}{\alpha}\hat{S}^\dag(r)\hat{R}^\dag(\theta)$. Action of this channel corresponds to symplectic matrix
\[
S(r,\theta)=R(\theta)S(r)=\begin{pmatrix}
e^{- i \theta}\cosh r & -e^{- i \theta}\sinh r \\
-e^{ i \theta}\sinh r & e^{ i \theta}\cosh r
\end{pmatrix},
\]
which leads to moments for $\R$,
\[
\bd=S(r,\theta)\bd_0
,
\quad \sigma=S(r,\theta)\sigma_0S^\dag(r,\theta).
\]
Since $\sigma$ is already written in the form of its symplectic decomposition, we can immediately use Eq.~\eqref{eq:multimode_QFI} to compute the QFIM. We compute $P_r=S^{-1}(r,\theta)\partial_r{S}(r,\theta)$, and $P_\theta=S^{-1}(r,\theta)\partial_\theta{S}(r,\theta)$, from which we obtain
\[
R_r=0,\ \,
Q_r=-1,\ \,
R_\theta=-i \cosh 2r,\ \,
Q_\theta=i \sinh 2r.
\]
Moreover,
\begin{align}
\sigma^{-1}&=\begin{pmatrix}
\cosh 2r & e^{-2 i \theta}\sinh 2r \\
e^{2 i \theta}\sinh 2r & \cosh 2r
\end{pmatrix},\nonumber\\
\partial_r\boldsymbol{d}&=\begin{pmatrix}
        e^{- i \theta}(-\ov{\alpha}\cosh r+\alpha \sinh r )\\
        e^{ i \theta}(-\alpha\cosh r+\ov{\alpha} \sinh r )\\
      \end{pmatrix},\\
\partial_\theta\boldsymbol{d}&=\begin{pmatrix}
        -ie^{- i \theta}(\alpha\cosh r-\ov{\alpha} \sinh r )\\
        ie^{ i \theta}(\ov{\alpha}\cosh r-\alpha \sinh r )\\
      \end{pmatrix}.\nonumber
\end{align}
\break
Since the symplectic eigenvalue $\lambda=1$, according to Eq.~\eqref{def:problematic_points}, the first and the third term in the sum, Eq.~\eqref{eq:multimode_QFI}, disappears, which yields
\[
H^{ij}(\be)=2\Re[\ov{Q_{i}}Q_{j}]
+2\partial_i\boldsymbol{d}^\dag\sigma^{-1}\partial_j\boldsymbol{d},
\]
from which we compute QFIM as
\begin{widetext}
\[
H(r,\theta)=
\begin{pmatrix}
         2+4\abs{\alpha}^2 & -4\Im[\alpha^2]\cosh 2r \\
         -4\Im[\alpha^2]\cosh 2r & 2\sinh^2 2r\!+\!4e^{4r}\Im[\alpha]^2\!+\!4e^{-4r}\Re[\alpha]^2 \\
       \end{pmatrix}.
\]
\end{widetext}

Further, applying $\lambda=1$ to Eq.~\eqref{eq:multimode_condition} we derive
\[
\tr[\R[\hat{\mathscr{L}}_i,\hat{\mathscr{L}}_j]]=4i\Im[\ov{Q_{i}}Q_{j}]+4\partial_i\bd^\dag\sigma^{-1}K\sigma^{-1}\partial_j\bd,
\]
which yields $\tr[\R[\hat{\mathscr{L}}_r,\hat{\mathscr{L}}_r]]=\tr[\R[\hat{\mathscr{L}}_\theta,\hat{\mathscr{L}}_\theta]]=0$, and
\[
\tr[\R[\hat{\mathscr{L}}_r,\hat{\mathscr{L}}_\theta]]=4i(-\sinh 2r+2e^{2r}\Im[\alpha]^2-2e^{-2r}\Re[\alpha]^2),
\]
which means that the quantum Cram\'er-Rao bound is not in general achievable for simultaneous estimation of $r$ and $\theta$ encoded into a coherent state.

\textit{\textbf{Limit formula}}

Here we are going to use Eqs.~\eqref{eq:limit_formula} and~\eqref{eq:remainder}  to numerically estimate QFIM of the Gaussian state from the first example, and then compared it to the analytical result.

From the first and the second moments, Eq.~\eqref{eq:first_example_moments}, we calculate
\[
\begin{split}
A&=K\sigma=\lambda\begin{pmatrix}
\cosh 2r & -\sinh 2r \\
\sinh 2r & -\cosh 2r
\end{pmatrix},\\
A^{-1}&=\frac{1}{\lambda}\begin{pmatrix}
\cosh 2r & -\sinh 2r \\
\sinh 2r & -\cosh 2r
\end{pmatrix},\\
\partial_\beta A&=\frac{\lambda^2-1}{2}\begin{pmatrix}
-\cosh 2r & \sinh 2r \\
-\sinh 2r & \cosh 2r
\end{pmatrix},\\
\partial_r A&=2\lambda\begin{pmatrix}
\sinh 2r & -\cosh 2r \\
\cosh 2r & -\sinh  2r
\end{pmatrix}.
\end{split}
\]
In order to calculate QFIM for $n$ decimal places,  we require $M$ to be such that
\[
|R_M^{\beta,\beta}|< \frac{1}{10^n},\quad |R_M^{\beta,r}|< \frac{1}{10^n},\quad|R_M^{rr}|< \frac{1}{10^n},
\]
which, using Eq.~\eqref{eq:remainder}, leads to
\[
M>\frac{n+\log_{10}\frac{\max_{i\in\{\beta,r\}}\mathrm{tr}[(A\partial_iA)^2]}{2(\lambda_{\mathrm{min}}^2-1)}}{2\log_{10}\lambda_{\min}}-1.
\]
To calculate QFIM for $\lambda=2$ and $r=1$ with precision for two decimal places, we insert $n=2$. In our example $\lambda_{\mathrm{min}}=\lambda$, which gives
\[
M>4.529,
\]
meaning that we need $M=5$ terms in the sum~\eqref{eq:limit_formula}. Summing these terms, we obtain an estimate for the QFIM,
\[
H(\beta,r)\approx
\begin{pmatrix}
         0.749268 & 0 \\
         0 & 3.20313 \\
       \end{pmatrix}.
\]
Comparing this to the analytical result calculated by inserting values $\lambda=2$ and $r=1$ into Eq.~\eqref{eq:example1exact},
\[
H(\beta,r)=
\begin{pmatrix}
         0.75 & 0 \\
         0 & 3.2 \\
       \end{pmatrix},
\]
shows that we are within the limit of two decimal places of precision.

\textit{\textbf{Pure states and discontinuity of QFIM}}

Here we show the difference between QFIM and cQFIM and show how it is connected to the discontinuity of the quantum Fisher information. We consider a task of estimating the squeezing parameter from the two-mode squeezed vacuum, $\R=\hat{S}_T(r)\pro{0}{0}\hat{S}_T^\dag(r)$, given by moments
\[\label{eq:full_state}
\bd=\begin{pmatrix}
        0 \\
        0 \\
        0 \\
        0 \\
      \end{pmatrix}
      ,
\quad \sigma=\begin{pmatrix}
\cosh 2r & 0  & 0 & -\sinh 2r \\
0 & \cosh 2r & -\sinh 2r & 0 \\
0 & -\sinh 2r & \cosh 2r & 0 \\
-\sinh 2r & 0 & 0 & \cosh 2r
\end{pmatrix}.
\]
Since this state is a pure state, we can use Eq.~\eqref{eq:pure_non-elegant} to compute the QFIM, and Eq.~\eqref{eq:pure_elegant} to compute the cQFIM. Variation of the parameter $r$ does not change the purity of the state, i.e., both symplectic eigenvalues $\lambda_1=\lambda_2=1$ for all $r$, thus, according to Eq.~\eqref{eq:QFIM_cQFIM_relation}, we should find $H=H_c$. Indeed, we have
\[
H(r)=H_c(r)=4.
\]

Now, let us say that experimenter does not have an access to the second mode, so they have to trace over it. In the phase-space formalism, this is done simply by taking out the rows and columns representing that mode~\cite{Weedbrook2012a} (in our notation, the second and the fourth row and column), resulting in state
\[\label{eq:reduced_state}
\bd_1=\begin{pmatrix}
        0 \\
        0 \\
      \end{pmatrix}
      ,
\quad \sigma_1=\begin{pmatrix}
\cosh 2r & 0  \\
0 & \cosh 2r
\end{pmatrix}.
\]
For $r=0$ the state is pure, so we can again use Eqs.~\eqref{eq:pure_non-elegant} and~\eqref{eq:pure_elegant} to compute QFIM and cQFIM. However, now we find
\[\label{eq:H_and_Hc}
H(0)=0,\quad H_c(0)=4.
\]
For $r>0$ the state is mixed, $\lambda_1=\cosh 2r$, according to Eq.~\eqref{eq:QFIM_cQFIM_relation}, $H=H_c$, and we can use any formula for mixed states (such as Eq.~\eqref{eq:multimode_QFI}) to compute the QFIM and cQFIM. Put together with Eq.~\eqref{eq:H_and_Hc}, we find
\[
H(r)=\begin{cases}
      0, & r=0, \\
      4, & \mathrm{otherwise},
   \end{cases}\quad
\quad H_c(r)=4.
\]
Clearly, QFIM is discontinuous at point $r=0$, as expected from the theory~\cite{vsafranek2017discontinuities}. Intuitively, this can be explained as follows: QFIM measures the amount of identifiability of parameter $dr$ from state $\R_{r+dr}$. As we can see from Eq.~\eqref{eq:reduced_state}, if $r=0$, then states $\R_{+dr}$ and $\R_{-dr}$ correspond to the same density matrix, $\R_{+dr}=\R_{-dr}$. Therefore, $dr$ is not identifiable around point $r=0$, because there is no physical experiment that experimenter could apply on the system to distinguish parameter $-dr$ from parameter $+dr$. It is therefore reasonable to expect that QFIM, which measures the ability to estimate $dr$, is zero at point $r=0$. Experimenter does not have the same problem when they have access to the full state, Eq.~\eqref{eq:full_state}, because in there $\R_{+dr}\neq\R_{-dr}$, so $dr$ is identifiable at point $r=0$. On the other hand, cQFIM, which is defined as four times the Bures metric, measures the infinitesimal distance between states $\R_{r}$ and $\R_{r+dr}$. This distance is always positive, no matter what $r$ is. Identifiability does not play any role in the measure of distance. It is therefore not surprising, that in this case, cQFIM is a continuous function in $r$.

\section{Conclusion}

In this paper we derived several expressions for the quantum Fisher information matrix (Eqs.~(\ref{eq:mixed_QFI},\ref{eq:multimode_QFI},\ref{eq:limit_formula},\ref{eq:pure_non-elegant})) for the multi-parameter estimation of multi-mode Gaussian states, associated symmetric logarithmic derivatives (Eqs.~(\ref{eq:SLD_mixed},\ref{eq:SLD_mixed_williamson},\ref{eq:sld_pure})), and expressions that determine saturability of the quantum Cram\'er-Rao bound (Eqs.~(\ref{eq:condition_QCR},\ref{eq:multimode_condition},\ref{eq:pure_state_condition})). We then illustrated their use on several examples.

As our main results, we consider expression for the QFIM when the Williamson's decomposition of the covariance matrix is known, Eq.~\eqref{eq:multimode_QFI}, which can be used for example for finding optimal Gaussian probe states for Gaussian unitary channels; the limit formula together with the estimate of the remainder, Eqs.~\eqref{eq:limit_formula} and \eqref{eq:remainder}, which can be used for efficient numerical calculations, to any given precision; and expressions for SLDs, Eqs.~(\ref{eq:SLD_mixed},\ref{eq:SLD_mixed_williamson},\ref{eq:sld_pure}), which can be studied to provide the optimal measurement schemes.

In addition, we discussed and resolved problematic behavior of QFIM at the points of purity, and we devised a regularization procedure (Eqs.~(\ref{eq:any_state_QFIM},\ref{eq:SLD_regularization},\ref{eq:condition_QCRregularization_procedure2})) that allows to use expressions for mixed states to calculate quantities for Gaussian states with any number of pure modes. Altogether, we provided a useful set of tools for Gaussian quantum metrology.

\textit{\textbf{Acknowledgements}} I thank Tanja Fabsits and Karishma Hathlia, for reading the first version of the manuscript, and for useful feedback. This research was supported by the Foundational Questions Institute (FQXi.org).

\appendix
\section{Derivation of formula for mixed states, the real form, and the symmetric logarithmic derivative.}\label{app:mixed_state}
Here we use the general result of~\cite{Gao2014a} to derive Eq.~\eqref{eq:mixed_QFI}. According to~\cite{Gao2014a} while using the Einstein's summation convention, the quantum Fisher information for $N$-mode Gaussian state can be calculated as
\[\label{eq:Gao_unpolished}
H_{i,j}(\be)=\frac{1}{2}(\mathfrak{M}^{-1})^{\alpha\beta,\mu\nu}\partial_j\Sigma^{\alpha\beta}\partial_i\Sigma^{\mu\nu}+(\Sigma^{-1})^{\mu\nu}\partial_j\boldsymbol{\lambda}^\mu\partial_i\boldsymbol{\lambda}^\nu.
\]
The displacement vector and the covariance matrix are defined as
$\boldsymbol{\lambda}^m=\mathrm{tr}\big[\hat{\rho}\boldsymbol{\hat{A}}_{G}^m\big]$ and $\Sigma^{mn}=\mathrm{tr}\big[\hat{\rho}\,\{(\boldsymbol{\hat{A}}_G-\boldsymbol{\lambda})^m,(\boldsymbol{\hat{A}}_G-\boldsymbol{\lambda})^n\}\big]$, $\bA_G=(\hat{a}_1,\hat{a}_1^\dag,\dots,\hat{a}_N,\hat{a}_N^\dag)^T$, and the symplectic form is given by $[\boldsymbol{\hat{A}}_{G}^m,\boldsymbol{\hat{A}}_{G}^{n}]=:\Omega^{mn}\mathrm{id}$. The inverse of the tensor $
\mathfrak{M}_G^{\alpha\beta,\mu\nu}=\Sigma^{\alpha\mu}\Sigma^{\beta\nu}+\frac{1}{4}\Omega^{\alpha\mu}\Omega^{\beta\nu}$ is defined via
\[\label{eq:inverse_matrix}
(\mathfrak{M}_G^{-1})^{\tilde{\mu}\tilde{\nu},\alpha\beta}\mathfrak{M}_G^{\alpha\beta,\mu\nu}=\delta^{\tilde{\mu}\mu}\delta^{\tilde{\nu}\nu},
\]
where $\delta^{\tilde{\mu}\mu}$ denotes the Kronecker delta. Considering the above definition, we can find a matrix form to Eq.~\eqref{eq:Gao_unpolished},
\[\label{eq:Gao_polished}
H^{ij}(\be)=\frac{1}{2}\vectorization{\partial_i\Sigma}^T\mathfrak{M}_G^{-1}\vectorization{\partial_j\Sigma}+\partial_i\boldsymbol{\lambda}^T\Sigma^{-1}\partial_j\boldsymbol{\lambda},
\]
where $\mathfrak{M}_G=\Sigma\otimes\Sigma+\frac{1}{4}\Omega\otimes\Omega$, $\otimes$ denotes the Kronecker product, and $\vectorization{\cdot}$ is a vectorization of a matrix.

To obtain the result in our notation we need to consider transformation relations
\[
\begin{split}
\sigma&=2P\Sigma XP^T,\\
K&=P\Omega XP^T,\\
\boldsymbol{d}&=P\boldsymbol{\lambda},
\end{split}
\]
where $X=\bigoplus_{i=1}^N\sigma_x$ ($X$ is real and $X^2=I$) and $P$ is a certain permutation matrix ($P$ is real and $PP^T=I$). Using properties
\[
\begin{split}
X\Sigma X&=\ov{\Sigma},\\
X\Omega X&=-\Omega,
\end{split}
\]
the fact that $\Omega$ is real, and identities
\[\label{id:Kronecker_product_ids}
\begin{split}
(ABC)\otimes(A'B'C')&=(A\otimes A')(B\otimes B')(C\otimes C'),\\
(C^T\otimes A)\vectorization{B}&=\vectorization{ABC},
\end{split}
\]
we derive
\[\label{eq:app_mixed_states}
H^{ij}(\be)=\frac{1}{2}\vectorization{\partial_i\sigma}^\dag\mathfrak{M}^{-1}\vectorization{\partial_j\sigma}+2\partial_i\boldsymbol{d}^\dag\sigma^{-1}\partial_j\boldsymbol{d},
\]
where $\mathfrak{M}=\ov{\sigma}\otimes\sigma-K\otimes K$.

Some authors also use the real form, which is defined with respect to the collection of quadrature operators $\boldsymbol{\hat{Q}}=\{\hat{x}_{1},\hat{x}_{2},\ldots,\hat{p}_{1},\hat{p}_{2},\ldots\}$, as $\bd_{R}^m=\mathrm{tr}\big[\hat{\rho}\boldsymbol{\hat{Q}}^m\big]$, $\sigma_{R}^{mn}=\mathrm{tr}\big[\hat{\rho}\{\Delta\boldsymbol{\hat{Q}}^m,\Delta\boldsymbol{\hat{Q}}^n\}\big]$, $\Delta\boldsymbol{\hat{Q}}:=\boldsymbol{\hat{Q}}-\bd_{R}$, $[\hat{{Q}}^{m},\hat{{Q}}^{n}]\,=i\,\Omega_R^{mn}\,\mathrm{id}$. Due to $\hat{a}_i=\frac{1}{\sqrt{2}}(\hat{x}_i+i\hat{p}_i)$ and $\hat{a}_i^\dag=\frac{1}{\sqrt{2}}(\hat{x}_i-i\hat{p}_i)$, the real form is connected to the complex form through a unitary matrix
\[\label{eq:U}
U=\frac{1}{\sqrt{2}}\,\begin{bmatrix}I & +iI \\ I & -iI\end{bmatrix}
\]
as 
\[\label{eq:matrix_transformation}
\bA=U\boldsymbol{\hat{Q}},\ \bd=U\bd_{R},\ \sigma=U\sigma_RU^\dag,\ \mathrm{and}\ K=Ui\Omega_RU^\dag.
\]
 Assuming the real, $\sigma_R=S_RD_RS_R^T$, and the complex, $\sigma=S D S^\dag$, symplectic decomposition, the diagonal matrix consisting of symplectic eigenvalues remains the same, $D_R=UDU^\dag=U^\dag DU=D$ which for symplectic matrices means that $\sigma=U\sigma_RU^\dag=US_RD_RS_R^T U^\dag=US_RU^\dag D U S_R^\dag U^\dag=US_RU^\dag D (US_RU^\dag)^\dag$ (where we have used that in the real form, the symplectic matrices are real, $S_R^T=S_R^\dag$), therefore
 \[\label{eq:symplectic_transformation}
 D=D_R,\ S=US_RU^\dag.
 \]
 
Using these transformation relations, Eqs.~\eqref{id:Kronecker_product_ids} and~\eqref{eq:app_mixed_states}, we derive
\[
H^{ij}(\be)=\frac{1}{2}\vectorization{\partial_i\sigma_R}^T\mathfrak{M}_R^{-1}\vectorization{\partial_j\sigma_R}+2\partial_i\boldsymbol{d}_R^T\sigma_R^{-1}\partial_j\boldsymbol{d}_R,
\]
where $\mathfrak{M}_R=\sigma_R\otimes\sigma_R-\Omega_R\otimes \Omega_R$.

Using a similar approach, we can rewrite expressions for the symmetric logarithmic derivatives originally published in Ref.~\cite{Gao2014a} in an elegant matrix form,
\[\label{eq:L}
\hat{\mathscr{L}}_i=\dbA^\dag\cA_i\dbA+2\dbA^\dag\sigma^{-1}\partial_i{\boldsymbol{d}}-\frac{1}{2}\tr[\sigma\cA_i],
\]
where $\dbA:=\bA-\bd$, $\vectorization{\cA_i}:=\mathfrak{M}^{-1}\vectorization{\partial_i\sigma}$. By $\dbA^\dag\cA_i\dbA$ we mean $ \sum_{m,n}\big(\dbA^m\big)^\dag\cA_i^{mn}\dbA^n$. $\cA_i$ is Hermitian, and is of form
\[\label{eq:structure_of_A}
\cA_i=\begin{bmatrix}
\cA_{Xi} & \cA_{Yi} \\
\overline{\cA_{Yi}} & \overline{\cA_{Xi}}
\end{bmatrix}.
\]
To show that, using $\vectorization{\cA_i}:=\mathfrak{M}^{-1}\vectorization{\partial_i\sigma}$ we can derive the defining equation for $\cA_i$,
\[
\sigma\cA_i\sigma-K\cA_iK=\partial_i\sigma.
\]
Taking the conjugate transpose of this equation, we find that if $\cA_i$ solves this equation, then also $\cA_i^\dag$ solves this equation. But since $\cA_i$ is uniquely defined when matrix $\mathfrak{M}$ is invertible (as $\vectorization{\cA_i}:=\mathfrak{M}^{-1}\vectorization{\partial_i\sigma}$), it must be that $\cA_i^\dag=\cA_i$, i.e., $\cA_i$ is Hermitian. Solution of the above equation can be also written as
\[\label{eq:cAlimit}
\cA_i=\sum_{n=1}^\infty(K\sigma)^{-n}K\partial_i\sigma K(\sigma K)^{-n}.
\]
This shows that $\cA_i$ is a combination of matrices $K$, $\sigma^{-1}=KSD^{-1}S^\dag K$, and $\sigma$. These matrices have the structure of Eq.~\eqref{eq:structure_of_A}, and because this structure is conserved when combining these matrices, also $\cA_i$ must have the same structure, i.e., Eq.~\eqref{eq:structure_of_A} holds.

\section{The phase-space representation of common Gaussian unitaries and Gaussian states}\label{sec:The_phase_space_Gaussian_unitaries}

Here we provide a list of complex and real symplectic matrices that represent often-used Gaussian unitaries (denoted with hat), and common Gaussian states. We use definitions introduced above Eq.~\eqref{def:first_and_second_moments} and Eq.~\eqref{eq:U}. The real form is denoted by the lower index $R$. This section is based on Chapter II~in Ref.~\cite{vsafranek2016gaussian} which contains more detailed discussion. The derivation uses Eq.~\eqref{def:transformation} and transformation relations~\eqref{eq:matrix_transformation} and~\eqref{eq:symplectic_transformation}.  One-mode operations acting on a multi-mode state (which leave the other modes invariant) are easily lifted into multi-mode operations by adding identities onto suitable places as illustrated on Eq.~\eqref{eq:phase_operator}.\\

\noindent
\emph{Rotation/phase-change} $\hat{R}(\theta)=\exp(-i\theta\hat{a}^\dag \hat{a})$, $\hat{R}_1(\theta)=\exp(-i\theta\hat{a}_1^\dag \hat{a}_1)$, is represented by the complex and real symplectic matrices
\begin{flalign}\label{eq:phase_operator}
R(\theta)&=\begin{bmatrix}
e^{-i\theta} & 0 \\
0 & e^{i\theta}
\end{bmatrix},
\quad R_R(\theta)=\begin{bmatrix}
\cos\theta & \sin\theta \\
-\sin\theta & \cos\theta
\end{bmatrix},&&\\
\quad R_1(\theta)&\!=\!\begin{bmatrix}
e^{-i\theta} & 0  & 0 & 0 \\
0 & 1 & 0 & 0 \\
0 & 0 & e^{i\theta} & 0 \\
0 & 0 & 0 & 1
\end{bmatrix},
\ 
R_{1R}(\theta)\!=\!\begin{bmatrix}
\cos\!\theta & 0  & \sin\!\theta & 0 \\
0 & 1 & 0 & 0 \\
-\sin\!\theta & 0 & \cos\!\theta & 0 \\
0 & 0 & 0 & 1
\end{bmatrix}.&&\nonumber
\end{flalign}
\emph{One-mode squeezing} $\hat{S}(r,\chi)=\exp(-\frac{r}{2}(e^{i\chi}\hat{a}^{\dag2}-e^{-i\chi}\hat{a}^{2}))$,
\begin{flalign}\label{eq:squeezing_operator}
S(r,\chi)&=\begin{bmatrix}
\cosh r & -e^{i\chi}\sinh r \\
-e^{-i\chi}\sinh r & \cosh r
\end{bmatrix},\\
\ S_R(r,\chi)&=\begin{bmatrix}
\cosh r-\cos\chi\sinh r & -\sin\chi\sinh r \\
-\sin\chi\sinh r & \cosh r+\cos\chi\sinh r
\end{bmatrix}.&&\nonumber
\end{flalign}
\emph{Mode-mixing} $\hat{B}(\theta,\chi)=\exp(\theta(e^{i\chi}\hat{a}_1^\dag\hat{a}_2-e^{-i\chi}\hat{a}_2^\dag\hat{a}_1))$,
\begin{flalign}\label{eq:mode_mixing_operator}
B(\theta,\chi)&\!=\!\begin{bmatrix}
\cos\theta & e^{i\chi}\sin\theta  & 0 & 0 \\
-e^{-i \chi}\sin\theta & \cos\theta & 0 & 0 \\
0 & 0 & \cos\theta & e^{-i \chi}\sin\theta \\
0 & 0 & -e^{i \chi}\sin\theta & \cos\theta
\end{bmatrix},&&\nonumber\\
B_R(\theta,\chi)&\!=\!\begin{bmatrix}
\cos\!\theta & \cos\!\chi\sin\!\theta & 0 & -\sin\!\chi\sin\!\theta \\
-\cos\chi\sin\!\theta & \cos\!\theta & -\sin\!\chi\sin\!\theta & 0 \\
0 & \sin\!\chi\sin\!\theta & \cos\!\theta & \cos\!\chi\sin\!\theta\\
\sin\!\chi\sin\!\theta & 0 & -\cos\!\chi\sin\!\theta & \cos\!\theta
\end{bmatrix}.&&
\end{flalign}
\emph{Two-mode squeezing} $\hat{S}_{T}(r,\chi)=\exp(-r(e^{i\chi}\hat{a}_1^\dag\hat{a}_2^\dag-e^{-i\chi}\hat{a}_1\hat{a}_2))$,
\begin{flalign}\label{eq:twomode_squeezing_operator}
&S_T(r,\chi)=\nonumber\\
&\begin{bmatrix}
\cosh r & 0  & 0 & -e^{i\chi}\sinh r \\
0 & \cosh r & -e^{i\chi}\sinh r & 0 \\
0 & -e^{-i\chi}\sinh r & \cosh r & 0 \\
-e^{-i\chi}\sinh r & 0 & 0 & \cosh r
\end{bmatrix},&&\nonumber\\
&S_{TR}(r,\chi)=\nonumber\\
&\begin{bmatrix}
\cosh r & -\cos \chi\sinh r  & 0 & -\sin \chi\sinh r \\
-\cos \chi\sinh r & \cosh r & -\sin \chi\sinh r & 0 \\
0 & -\sin \chi\sinh r & \cosh r & \cos \chi\sinh r \\
-\sin \chi\sinh r & 0 & \cos \chi\sinh r & \cosh r
\end{bmatrix}.&&
\end{flalign}

Sometimes in the literature, yet another notation is encountered, for example it is common to see the real form displacement vectors and covariance matrices generated by ordering `$xpxp$' given by $\boldsymbol{\hat{Q}}:=(\hat{x}_1,\hat{p}_1,\hat{x}_2,\hat{p}_2)^T$ instead of `$xxpp$' vector $\boldsymbol{\hat{Q}}:=(\hat{x}_1,\hat{x}_2,\hat{p}_1,\hat{p}_2)^T$ used here.  It is easy to transform into the `$xpxp$' form by simply reordering rows and columns,
\[
\begin{split}
S_R&=\begin{bmatrix}
S_{x_1x_1} & S_{x_1x_2}  & S_{x_1p_1} & S_{x_1p_2} \\
S_{x_2x_1} & S_{x_2x_2} & S_{x_2p_1} & S_{x_2p_2} \\
S_{p_1x_1} & S_{p_1x_2} & S_{p_1p_1} & S_{p_1p_2} \\
S_{p_2x_1} & S_{p_2x_2} & S_{p_2p_1} & S_{p_2p_2}
\end{bmatrix}\\ 
&\longrightarrow\\
S_{R,xpxp}&=\begin{bmatrix}
S_{x_1x_1} & S_{x_1p_1}  & S_{x_1x_2} & S_{x_1p_2} \\
S_{p_1x_1} & S_{p_1p_1} & S_{p_1x_2} & S_{p_1p_2} \\
S_{x_2x_1} & S_{x_2p_1} & S_{x_2x_2} & S_{x_2p_2} \\
S_{p_2x_1} & S_{p_2p_1} & S_{p_2x_2} & S_{p_2p_2}
\end{bmatrix},
\end{split}
\]
which corresponds to transformation with a permutation matrix,
\[
S_{R,xpxp}=PS_{R}P^T,\ P=\begin{bmatrix}
1 & 0  & 0 & 0 \\
0 & 0 & 1 & 0 \\
0 & 1 & 0 & 0 \\
0 & 0 & 0 & 1
\end{bmatrix}.
\]

Now we introduce the most common Gaussian states. Characteristics of all other Gaussian states can be seen as mixtures of characteristics of these basic ones. In that sense the following list is complete.

\emph{Thermal state} is the simplest Gaussian state. Assuming the single particle Hilbert space is spanned by $N$ states -- modes, each mode is characterized by the energy $E_i$ of the state $\ket{\psi_i}$. We assume that each mode is thermally populated, i.e., number of particles in each mode is given by the thermal distribution, $\R_{{\mathrm{th}}i}=\frac{1}{Z}\mathrm{exp}(-\frac{E_i}{kT}\hat{n}_i)$, where $\hat{n}_i=\hat{a}_i^\dag\hat{a}_i$ denotes the number operator associated with mode $i$, $k$ is the Boltzmann constant, and $Z=\tr[e^{-\frac{E_i}{kT}\hat{n}_i}]$ defines the partition function. 
The full thermal state is then a tensor product of the thermal states of each mode, $\R_{\mathrm{th}}=\hat{\rho}_{\mathrm{th}1}\otimes\cdots\otimes\hat{\rho}_{\mathrm{th}N}$. The displacement vector of the thermal state is equal to zero and the covariance matrix in both complex and the real form is a diagonal matrix,
\begin{align}
\bd&=\boldsymbol{0}, \quad \sigma=\mathrm{diag}(\lambda_1,\dots,\lambda_N,\lambda_1,\dots,\lambda_N),\\
\bd_R&=\boldsymbol{0}, \quad \sigma_{R}=\mathrm{diag}(\lambda_1,\dots,\lambda_N,\lambda_1,\dots,\lambda_N),
\end{align}
where $\lambda_i=\coth(\frac{E_i}{2kT})$ are the symplectic eigenvalues. They can be also expressed in terms of the mean number of thermal bosons, $\lambda_i=1+2n_{{\mathrm{th}}i}$, where $n_{{\mathrm{th}}i}:=\tr[\hat{n}_i\R_{\mathrm{th}}]$.
Larger temperatures and smaller energies correspond to larger symplectic eigenvalues. For each $i$, $\lambda_i\geq1$ and $\lambda_i=1$ for $T=0$. Thermal state corresponding to $T=0$ is the lowest-energy state called \emph{vacuum state} and is described by the identity matrix $\sigma=I$.

\emph{Coherent state} is a Gaussian state which is characterized only by its displacement vector,
\[
\ket{\A}=e^{-\frac{\abs{\A}^2}{2}}\sum_{n=0}^\infty \frac{\A^n}{\sqrt{n!}}\ket{n}.
\]
Coherent state is an eigenvector of the annihilation operator, $a\ket{\A}=\A\ket{\A}$. Coherent states typically describe beams of light emitted by a laser~\cite{zhang1990coherent}. Mathematically, coherent state can be created by the action of the Weyl displacement operator on the vacuum (thus an equivalent name would be a single-mode displaced vacuum), $\ket{\A}=\hat{D}(\A)\ket{0}$. The first and the second moments are
\begin{align}
\bd&=(\A,\ov{\A})^T, \quad \sigma=I,\\
\bd_R&=\sqrt{2}(\Re[\A],\Im[\A])^T, \quad \sigma_R=I.
\end{align}

\emph{(Single-mode) squeezed state} is created by an action of the squeezing operator~\eqref{eq:squeezing_operator} on the vacuum, $\ket{S(r,\chi)}=\hat{S}(r,\chi)\ket{0}$. For $\chi=0$ this state takes the form~\cite{kok2010introduction}
\[
\ket{S(r)}=\frac{1}{\sqrt{\cosh\abs{r}}}\sum_{n=0}^\infty\frac{\sqrt{(2n)!}}{n!}\left(\frac{-r}{2\abs{r}}\right)^n\tanh^n\abs{r}\ket{2n}.
\]
Such states for example from a laser light by going through an optical parametric oscillator~\cite{breitenbach1997measurement,Lvovsky2014squeezed}.
The first and the second moments are
\begin{align}
\bd&=\boldsymbol{0}, \quad \sigma=S(r,\chi)S^\dag(r,\chi)=S(2r,\chi),\\
\bd_R&=\boldsymbol{0}, \quad \sigma_R=S_R(r,\chi)S_R^T(r,\chi)=S_R(2r,\chi).
\end{align}

\emph{Two-mode squeezed state} is an entangled two-mode state created by an action of the two mode squeezing operator~\eqref{eq:twomode_squeezing_operator} on the vacuum, $\ket{S_T(r,\chi)}=\hat{S}_T(r,\chi)\ket{0}$. For $\chi=0$ this state takes the form~\cite{kok2010introduction}
\[
\ket{S_T(r)}=\frac{1}{\cosh\abs{r}}\sum_{n=0}^\infty\left(\frac{-r}{\abs{r}}\right)^n\tanh^n\abs{r}\ket{n,n}
\]
Physically, two-mode squeezed states are prepared by sending squeezed and anti-squeezed state (squeezed with the negative squeezing) through a beam-splitter (mode-mixing operator \eqref{eq:mode_mixing_operator}  with $\chi=0$). The first and the second moments are
\begin{align}
\bd&=\boldsymbol{0}, \quad \sigma=S_T(r,\chi)S_T^\dag(r,\chi)=S_T(2r,\chi),\\
\bd_R&=\boldsymbol{0}, \quad \sigma_R=S_{TR}(r,\chi)S_{TR}^T(r,\chi)=S_{TR}(2r,\chi).
\end{align}

\section{Saturability of the Cram\'er-Rao bound}\label{app:saturability}
Here we derive expression for $\tr[\R[\hat{\mathscr{L}}_i,\hat{\mathscr{L}}_j]]$ in the phase-space formalism, which gives condition on the saturability of the quantum Cram\'er-Rao bound, Eq.~\eqref{eq:condition_for_CRbound}.

Inserting Eq.~\eqref{eq:L} we derive
\[\label{eq:condition_wide}
\begin{split}
\tr[\R[\hat{\mathscr{L}}_i,\hat{\mathscr{L}}_j]]&=\tr[\R[\dbA^\dag\cA_i\dbA,\dbA^\dag\cA_j\dbA]]\\
&+\tr[\R[\dbA^\dag\cA_i\dbA,2\dbA^\dag\sigma^{-1}\partial_j{\boldsymbol{d}}]]\\
&+\tr[\R[2\dbA^\dag\sigma^{-1}\partial_i{\boldsymbol{d}},\dbA^\dag\cA_i\dbA]]\\
&+\tr[\R[2\dbA^\dag\sigma^{-1}\partial_i{\boldsymbol{d}},2\dbA^\dag\sigma^{-1}\partial_j{\boldsymbol{d}}]]\\
&\equiv a_1+a_2+a_3+a_4
\end{split}
\]
Terms that contain $\frac{1}{2}\tr[\sigma\cA_i]$ vanish, because as a number, this term commutes with every operator.

In the following, we use the Einstein's summation convention. We will use the commutation relations,
\begin{subequations}
\begin{align}
[\dbA^m,\dbA^{\dag n}]&=K^{mn}\mathrm{id},\\
[\dbA^{m},\dbA^{n}]&=(KT)^{mn}\mathrm{id},\\
[\dbA^{\dag m},\dbA^{\dag n}]&=(TK)^{mn}\mathrm{id}=-(KT)^{nm}\mathrm{id},
\end{align}
\end{subequations}
and identities,
\begin{subequations}
\begin{align}
\dbA^{\dag m}\dbA^{n}&=\tfrac{1}{2}(\{\dbA^{n},\dbA^{\dag m}\}-K^{nm}\mathrm{id}),\\
\dbA^{m}\dbA^{n}&=\tfrac{1}{2}(\{\dbA^{n},\dbA^{\dag k}\}-K^{nk}\mathrm{id})T^{km},\\
\dbA^{\dag m}\dbA^{\dag n}&=\tfrac{1}{2}T^{nk}(\{\dbA^{k},\dbA^{\dag m}\}-K^{km}\mathrm{id}),
\end{align}
\end{subequations}
from which follows
\begin{subequations}
\begin{align}
\tr[\R\dbA^{\dag m}\dbA^{n}]&=\tfrac{1}{2}(\sigma-K)^{nm},\\
\tr[\R\dbA^{m}\dbA^{n}]&=\tfrac{1}{2}(\sigma-K)^{nk}T^{km},\\
\tr[\R\dbA^{\dag m}\dbA^{\dag n}]&=\tfrac{1}{2}T^{nk}(\sigma-K)^{km}.
\end{align}
\end{subequations}
We have used $\dbA^\dag=T\dbA$, where matrix $T$ is defined as
\[\label{def:T}
T=
\begin{bmatrix}
0 & I \\
I & 0
\end{bmatrix}.
\]
Further, we use properties of commutator,
\begin{subequations}
\begin{align}
[AB,C]&=A[B,C]+[A,C]B,\\
[A,BC]&=B[A,C]+[A,B]C.
\end{align}
\end{subequations}
We look separately at each term $a_1,a_2,a_3,a_4$.

\begin{widetext}
\[
\begin{split}
a_1&=\cA_i^{kl}\cA_j^{mn}\tr[\R[\dbA^{k\dag}\dbA^l,\dbA^{m\dag}\dbA^n]]\\
&=\cA_i^{kl}\cA_j^{mn}\tr[\R(\dbA^{k\dag}\dbA^{m\dag}[\dbA^l,\dbA^n]+\dbA^{k\dag}[\dbA^l,\dbA^{m\dag}]\dbA^n\\
&\ \ \ \ \ \ \ \ \ \ \ \ \ \ \ \ \ \ \ +\dbA^{m\dag}[\dbA^{k\dag},\dbA^n]\dbA^l+[\dbA^{k\dag},\dbA^{m\dag}]\dbA^n\dbA^l)]\\
&=\cA_i^{kl}\cA_j^{mn}\tr[\R(\dbA^{k\dag}\dbA^{m\dag}(KT)^{ln}+\dbA^{k\dag}K^{lm}\dbA^n-\dbA^{m\dag}K^{kn}\dbA^l-(KT)^{mk}\dbA^n\dbA^l)]\\
&=\tfrac{1}{2}\cA_i^{kl}\cA_j^{mn}(T^{ms}(\sigma-K)^{sk}(KT)^{ln}+K^{lm}(\sigma-K)^{nk}-K^{kn}(\sigma-K)^{lm}-(KT)^{km}(\sigma-K)^{ls}T^{sn})]\\
&=\tfrac{1}{2}\big(T^{ms}(\sigma-K)^{sk}\cA_i^{kl}(KT)^{ln}(\cA_j^T)^{nm}+\cA_i^{kl}K^{lm}\cA_j^{mn}(\sigma-K)^{nk}\\
&\ \ \ -K^{nk}\cA_i^{kl}(\sigma-K)^{lm}\cA_j^{mn}-(TK)^{mk}\cA_i^{kl}(\sigma-K)^{ls}T^{sn}(\cA_j^T)^{mn}\big)\\
&=\tfrac{1}{2}\tr[(K\sigma-I)(\cA_iK(T\cA_j^TT)K+\cA_iK\cA_jK-\cA_jK\cA_iK-(T\cA_j^TT)K\cA_iK)]\\
&=\tr[(K\sigma-I)[\cA_iK,\cA_jK]]\\
&=\tr[K\sigma[\cA_iK,\cA_jK]],
\end{split}
\]
where we have used $T\cA_j^TT=\cA_j$, which holds for any matrix of form Eq.~\eqref{eq:structure_of_A}, and the fact that trace of a commutator is always zero.

Using the properties of vectorization, Eq.~\eqref{id:Kronecker_product_ids}, the fact that $\cA_i$ and $\mathfrak{M}$ are Hermitian, and $\vectorization{\cA_i}=\mathfrak{M}^{-1}\vectorization{\partial_i\sigma}$, we further derive
\[\label{eq:first_part}
\begin{split}
a_1&=\tr[\cA_i^\dag K\cA_j\sigma-\cA_i^\dag\sigma\cA_jK]\\
&=\vectorization{\cA_i}^\dag(\ov{\sigma}\!\otimes\! K\!-\!K\!\otimes\!\sigma)\vectorization{\cA_j}\\
&=\vectorization{\partial_i\sigma}^\dag\mathfrak{M}^{-1}(\ov{\sigma}\!\otimes\! K\!-\!K\!\otimes\!\sigma)\mathfrak{M}^{-1}\vectorization{\partial_j\sigma}.
\end{split}
\]

Terms $a_2$ and $a_3$ in Eq.~\eqref{eq:condition_wide} are identically zero, because contracting a commutation relation between two operators will always give a number, and the remaining operator commutes with a number.

Finally, we have
\[\label{eq:second_part}
\begin{split}
a_4&=4(\sigma^{-1})^{kl}(\sigma^{-1})^{mn}\partial_j{\boldsymbol{d}}^n\partial_i{\boldsymbol{d}}^l\tr[\R[\dbA^{k\dag},\dbA^{m\dag}]]\\
&=4(\sigma^{-1})^{kl}(\sigma^{-1})^{mn}\partial_j{\boldsymbol{d}}^nT^{ls}\ov{\partial_i{\boldsymbol{d}}^s}\tr[\R(TK)^{km}]\\
&=4\ov{\partial_i{\boldsymbol{d}}^s}T^{sl}((\sigma^{-1})^T)^{lk}(TK)^{km}(\sigma^{-1})^{mn}\partial_j{\boldsymbol{d}}^n\\
&=4\partial_i{\boldsymbol{d}}^\dag T((\sigma^{-1})^T)TK\sigma^{-1}\partial_j{\boldsymbol{d}}\\
&=4\partial_i{\boldsymbol{d}}^\dag \sigma^{-1}K\sigma^{-1}\partial_j{\boldsymbol{d}},
\end{split}
\]
where we have used $T((\sigma^{-1})^T)T=\sigma^{-1}$, because $\sigma^{-1}$ has the same structure as Eq.~\eqref{eq:structure_of_A}. (This can be for example seen from the Williamson's decomposition $\sigma^{-1}=KSD^{-1}S^\dag K$, where each of the matrices $K,S,D^{-1},...$ has this structure, therefore also $\sigma^{-1}$ has that structure.)

Combining Eq.~\eqref{eq:condition_wide}, \eqref{eq:first_part}, and \eqref{eq:second_part}, we derive
\[
\tr[\R[\hat{\mathscr{L}}_i,\hat{\mathscr{L}}_j]]=\vectorization{\partial_i\sigma}^\dag\mathfrak{M}^{-1}(\ov{\sigma}\!\otimes\! K\!-\!K\!\otimes\!\sigma)\mathfrak{M}^{-1}\vectorization{\partial_j\sigma}+4\partial_i{\boldsymbol{d}}^\dag \sigma^{-1}K\sigma^{-1}\partial_j{\boldsymbol{d}}.
\]
\end{widetext}

\section{Finding the Williamson's decomposition of the covariance matrix}\label{app:diagonalization}

Here we show how to find the symplectic matrices that diagonalize the covariance matrix, $\sigma=SDS^\dag$, where $D=\mathrm{diag}(\lambda_1,\dots,\lambda_N,\lambda_1,\dots,\lambda_N)$ is a diagonal matrix consisting of \emph{symplectic eigenvalues}, which are defined as the positive eigenvalues of the matrix $K\sigma$. We note, that eigenvalues of $K\sigma$ always come in pairs~\cite{deGosson2006a,Adesso2014a}, i.e., when the $\lambda_i$ is an eigenvalue of $K\sigma$, then $-\lambda_i$ is also an eigenvalue of $K\sigma$. As a result, $KD=\mathrm{diag}(\lambda_1,\dots,\lambda_N,-\lambda_1,\dots,-\lambda_N)$ is a diagonal matrix that consists of the entire spectrum of matrix $K\sigma$.

Essentially, we will redo part of the proof of the Williamson's theorem (see, e.g., Ref.~\cite{simon1999congruences}), but in the complex form of the density matrix, where the symplectic matrices are defined by the complex form of the real symplectic group, Eq.~\eqref{def:structure_of_S}. The reader is welcome to skip the derivation, and go directly to the summarized result, Theorem~\ref{thm:Williamson}.

We would like to find symplectic matrices such that
\[\label{eq:decomposition_sigma}
\sigma=SDS^\dag.
\]
It can be easily checked that
\[\label{eq:tildeS}
S=\sigma^{\frac{1}{2}}UD^{-\frac{1}{2}}
\]
where $U$ is any unitary matrix, solves Eq.~\eqref{eq:decomposition_sigma}. $\sigma^{\frac{1}{2}}$ always exists, because $\sigma$ is a positive-definite matrix. For generic $U$, this solution may not be a symplectic matrix, but we will manage to find $U$ so that $S$ is indeed symplectic. To do that, we plug this solution into the defining relation for the symplectic matrix, Eq.~\eqref{def:structure_of_S}, and obtain
\[
\sigma^{\frac{1}{2}}UD^{-\frac{1}{2}}KD^{-\frac{1}{2}}U^\dag\sigma^{\frac{1}{2}}=K.
\]
We invert both sides, and rewrite it as
\[\label{eq:KD}
KD=U^\dag\sigma^{\frac{1}{2}}K\sigma^{\frac{1}{2}}U.
\]
On the left hand side, we have a diagonal matrix consisting of eigenvalues of $K\sigma$. According to this equation, unitary matrix $U$ should be the matrix that diagonalizes matrix $\sigma^{\frac{1}{2}}K\sigma^{\frac{1}{2}}$. But does such $U$ exist? Such $U$ exists if and only if eigenvalues of $K\sigma$ and $\sigma^{\frac{1}{2}}K\sigma^{\frac{1}{2}}$ are identical. However, as we can easily see from the characteristic polynomial,
\[
\begin{split}
|\sigma^{\frac{1}{2}}K\sigma^{\frac{1}{2}}-\lambda|&=|\sigma^{\frac{1}{2}}||K\sigma^{\frac{1}{2}}-\lambda\sigma^{-\frac{1}{2}}|\\
&=|K\sigma^{\frac{1}{2}}-\lambda\sigma^{-\frac{1}{2}}||\sigma^{\frac{1}{2}}|=|K\sigma-\lambda|,
\end{split}
\]
this is indeed the case. Therefore, we have found a choice of $U$ that will make sure that $S$ is symplectic. We can summarize our findings in the following Theorem:

\begin{theorem}\label{thm:Williamson}(Williamson)
Any covariance matrix can be decomposed using symplectic matrices as $\sigma=SDS^\dag$.
Symplectic matrix $S$ is calculated as
\[
S=\sigma^{\frac{1}{2}}UD^{-\frac{1}{2}},
\]
where unitary matrix $U$ consists of eigenvectors of $\sigma^{\frac{1}{2}}K\sigma^{\frac{1}{2}}$, i.e., it solves Eq.
\[
U^\dag\sigma^{\frac{1}{2}}K\sigma^{\frac{1}{2}}U=KD,
\]
and $D=\mathrm{diag}(\lambda_1,\dots,\lambda_N,\lambda_1,\dots,\lambda_N)$, where $\lambda_i$ are the positive eigenvalues of operator $\sigma^{\frac{1}{2}}K\sigma^{\frac{1}{2}}$ (or equivalently, of operator $K\sigma$).
\end{theorem}

\section{Derivation of formulas for the case when the Williamson's decomposition is known}\label{app:williamson_QFI}
Here we use Eq.~\eqref{eq:mixed_QFI} to derive Eq.~\eqref{eq:multimode_QFI}. Using the Williamson decomposition $\sigma=SDS^\dag$, identities \eqref{def:structure_of_S} and \eqref{id:Kronecker_product_ids} we derive
\[
\begin{split}
\mathfrak{M}^{-1}&=(\!(S^{-1}\!)^T\!\otimes\! (\!KSK\!)\!)(\!D\!\otimes\! D\!-\!K\!\otimes\! K)^{-1}(\!(\!KSK\!)^T\!\otimes\! S^{-1}\!),\\
\partial_i\sigma&=\partial_iS DKS^{-1}K+S\partial_iD KS^{-1}K-SDKS^{-1}\partial_iSS^{-1}K.
\end{split}
\]
The first part of Eq.~\eqref{eq:mixed_QFI} then reads
\begin{multline}\label{eq:P1_first_part}
\frac{1}{2}\vectorization{\partial_i\sigma}^\dag\mathfrak{M}^{-1}\vectorization{\partial_j\sigma}=\\
\vectorization{\!P_{i}D\!-\!DKP_{i}K\!+\!\partial_iD\!}^\dag\mathfrak{M}_{diag}^{-1}\vectorization{\!P_{j}D\!-\!DKP_{j}K\!+\!\partial_jD\!},
\end{multline}
where $\mathfrak{M}_{diag}=D\!\otimes\! D\!-\!K\!\otimes \!K$.

We define elements of matrix $X$ in basis $\ket{\mu}\ket{\nu}$ as
\[
X=\sum_{\mu,\nu,\tilde{\mu},\tilde{\nu}}X^{\tilde{\mu}\tilde{\nu},\mu\nu}\ket{\tilde{\mu}}\ket{\tilde{\nu}}\bra{\mu}\bra{\nu},
\]
and applying definition of vectorization, Eq.~\eqref{eq:vectorizationdef}, on matrices $Y=\sum_{\alpha,\beta}Y^{\alpha\beta}\ket{\alpha}\bra{\beta}$, $Z=\sum_{\alpha,\beta}Z^{\alpha\beta}\ket{\alpha}\bra{\beta}$, we derive
\begin{subequations}
\begin{align}
\vectorization{Z}&=\sum_{\alpha,\beta}Z^{\alpha\beta}\ket{\beta}\ket{\alpha},\\
\vectorization{Y}^\dag&=\sum_{\alpha,\beta}\ov{Y}^{\alpha\beta}\bra{\beta}\bra{\alpha}.
\end{align}
\end{subequations}
Then we have
\[\label{eq:vecyXz}
\begin{split}
&\vectorization{Y}^\dag X \vectorization{Z}\\
&=\!\!\!\sum_{\alpha,\beta,\tilde{\alpha},\tilde{\beta},\mu,\nu,\tilde{\mu},\tilde{\nu}}\!\!\!\ov{Y}^{\tilde{\alpha}\tilde{\beta}}\bra{\tilde{\beta}}\bra{\tilde{\alpha}} \bigg(X^{\tilde{\mu}\tilde{\nu},\mu\nu}\ket{\tilde{\mu}}\ket{\tilde{\nu}}\bra{\mu}\bra{\nu}\bigg) Z^{\alpha\beta}\ket{\beta}\ket{\alpha}\\
&=\!\!\!\sum_{\alpha,\beta,\tilde{\alpha},\tilde{\beta},\mu,\nu,\tilde{\mu},\tilde{\nu}}\!\!\!\ov{Y}^{\tilde{\alpha}\tilde{\beta}}X^{\tilde{\mu}\tilde{\nu},\mu\nu} Z^{\alpha\beta}
\delta^{\tilde{\beta}\tilde{\mu}}\delta^{\tilde{\alpha}\tilde{\nu}}\delta^{{\beta}{\mu}}\delta^{{\alpha}{\nu}}\\
&=\!\!\!\sum_{\mu,\nu,\tilde{\mu},\tilde{\nu}}\!\!\!\ov{Y}^{\tilde{\nu}\tilde{\mu}}X^{\tilde{\mu}\tilde{\nu},\mu\nu} Z^{\nu\mu}
\end{split}
\]
Similarly, defining matrix $W$ such that
\[
\vectorization{W}=X \vectorization{Z}=\sum_{\mu,\nu}X^{\tilde{\mu}\tilde{\nu},\mu\nu} Z^{\nu\mu}\ket{\tilde{\mu}}\ket{\tilde{\nu}},
\]
we find that this matrix has elements
\[\label{eq:vecW}
W^{\tilde{\nu}\tilde{\mu}}=\sum_{\mu,\nu}X^{\tilde{\mu}\tilde{\nu},\mu\nu} Z^{\nu\mu}.
\]

In this formalism, since
\[
\begin{split}
\mathfrak{M}_{diag}&=(\sum_{\tilde{\mu},\mu}D^{\tilde{\mu}\mu}\ket{\tilde{\mu}}\bra{\mu})\otimes(\sum_{\tilde{\nu},\nu}D^{\tilde{\nu}\nu}\ket{\tilde{\nu}}\bra{\nu})\\
&-(\sum_{\tilde{\mu},\mu}K^{\tilde{\mu}\mu}\ket{\tilde{\mu}}\bra{\mu})\otimes(\sum_{\tilde{\nu},\nu}K^{\tilde{\nu}\nu}\ket{\tilde{\nu}}\bra{\nu})\\
&=\sum_{\mu,\nu,\tilde{\mu},\tilde{\nu}}(D^{\tilde{\mu}\mu}D^{\tilde{\nu}\nu}-K^{\tilde{\mu}\mu}K^{\tilde{\nu}\nu})\ket{\tilde{\mu}}\ket{\tilde{\nu}}\bra{\mu}\bra{\nu},
\end{split}
\] and $D^{\tilde{\mu}\mu}=\delta^{\tilde{\mu}\mu} D^{\mu\mu}$, $K^{\tilde{\mu}\mu}=\delta^{\tilde{\mu}\mu} K^{\mu\mu}$, we find
\[
(\mathfrak{M}_{diag}^{-1})^{\tilde{\mu}\tilde{\nu},\mu\nu}=\frac{\delta^{\tilde{\mu}\mu}\delta^{\tilde{\nu}\nu}}{D^{\mu\mu}D^{\nu\nu}-K^{\mu\mu}K^{\nu\nu}}.
\]
Changing to element-wise notation and using Einstein's summation convention ($\mu,\nu\in\{1,\dots,2N\}$, $k,l\in\{1,\dots,N\}$) we expand Eq.~\eqref{eq:P1_first_part} using Eq.~\eqref{eq:vecyXz},
\begin{widetext}
\[
\begin{split}
&\frac{1}{2}\vectorization{\partial_i\sigma}^\dag\mathfrak{M}^{-1}\vectorization{\partial_j\sigma}=\frac{1}{2}(\ov{P_{i}}D-DK\ov{P_{i}}K+\partial_iD)^{\tilde{\nu}\tilde{\mu}}
\frac{\delta^{\tilde{\mu}\mu}\delta^{\tilde{\nu}\nu}}{D^{\mu\mu}D^{\nu\nu}-K^{\mu\mu}K^{\nu\nu}}
(P_{j}D-DKP_{j}K+\partial_jD)^{\nu\mu}\\
&=\frac{1}{2}
\frac{(\ov{P_{i}}^{\nu\mu}D^{\mu\mu}-D^{\nu\nu} K^{\nu\nu}\ov{P_{i}}^{\nu\mu}K^{\mu\mu}+\partial_iD^{\nu\nu}\delta^{\nu\mu})(P_{j}^{\nu\mu}D^{\mu\mu}-D^{\nu\nu} K^{\nu\nu} P_{j}^{\nu\mu}K^{\mu\mu}+\partial_jD^{\nu\nu}\delta^{\nu\mu})}{D^{\nu\nu}D^{\mu\mu}-K^{\nu\nu}K^{\mu\mu}}\\
&=\frac{1}{2}\bigg(
\frac{1}{\lambda_k\lambda_l-1}
(\ov{R_{i}}^{kl}\lambda_l-\lambda_k\ov{R_{i}}^{kl}+\partial_i\lambda_k\delta^{kl})(R_{j}^{kl}\lambda_l-\lambda_k R_{j}^{kl}+\partial_j\lambda_k\delta^{kl})\\
&+\frac{1}{\lambda_k\lambda_l-1}
({R_{i}}^{kl}\lambda_l-\lambda_k{R_{i}}^{kl}+\partial_i\lambda_k\delta^{kl})(\ov{R_{j}}^{kl}\lambda_l-\lambda_k \ov{R_{j}}^{kl}+\partial_j\lambda_k\delta^{kl})\\
&+\frac{1}{\lambda_k\lambda_l+1}(\ov{Q_{i}}^{kl}\lambda_l+\lambda_k \ov{Q_{i}}^{kl})(Q_{j}^{kl}\lambda_l+\lambda_k Q_{j}^{kl})
+\frac{1}{\lambda_k\lambda_l+1}(Q_{i}^{kl}\lambda_l+\lambda_k Q_{i}^{kl})(\ov{Q_{j}}^{kl}\lambda_l+\lambda_k \ov{Q_{j}}^{kl})\bigg)\\
&=\frac{1}{2}\bigg(
\frac{1}{\lambda_k\lambda_l-1}
((\lambda_l-\lambda_k)\ov{R_{i}}^{kl}+\partial_i\lambda_k\delta^{kl})((\lambda_l-\lambda_k)R_{j}^{kl}+\partial_j\lambda_k\delta^{kl})+\frac{1}{\lambda_k\lambda_l+1}
(\lambda_l+\lambda_k)^2\ov{Q_{i}}^{kl}Q_{j}^{kl}\\
&+\frac{1}{\lambda_k\lambda_l-1}
((\lambda_l-\lambda_k){R_{i}}^{kl}+\partial_i\lambda_k\delta^{kl})((\lambda_l-\lambda_k)\ov{R_{j}}^{kl}+\partial_j\lambda_k\delta^{kl})+\frac{1}{\lambda_k\lambda_l+1}
(\lambda_l+\lambda_k)^2Q_{i}^{kl}\ov{Q_{j}}^{kl}\bigg)\\
&=\frac{(\lambda_l-\lambda_k)^2}{\lambda_k\lambda_l-1}\Re[\ov{R_{i}}^{kl}R_{j}^{kl}]+\frac{(\lambda_l+\lambda_k)^2}{\lambda_k\lambda_l+1}\Re[\ov{Q_{i}}^{kl}Q_{j}^{kl}]
+\frac{\partial_i\lambda_k\partial_j\lambda_k}{\lambda_k^2-1},
\end{split}
\]
which in combination with term $2\partial_i\boldsymbol{d}^\dag\sigma^{-1}\partial_j\boldsymbol{d}$ gives Eq.~\eqref{eq:multimode_QFI}.

Similarly, we derive
\[
\begin{split}
&\vectorization{\partial_i\sigma}^\dag\mathfrak{M}^{-1}(\ov{\sigma}\!\otimes\! K\!-\!K\!\otimes\!\sigma)\mathfrak{M}^{-1}\vectorization{\partial_j\sigma}\\
&=\vectorization{\!P_{i}D\!-\!DKP_{i}K\!+\!\partial_iD\!}^\dag\mathfrak{M}_{diag}^{-1}(D\!\otimes\! K\!-\!K\!\otimes\!D)\mathfrak{M}_{diag}^{-1}\vectorization{\!P_{j}D\!-\!DKP_{j}K\!+\!\partial_jD\!},\\
&=(\ov{P_{i}}D-DK\ov{P_{i}}K+\partial_iD)^{\tilde{\nu}\tilde{\mu}}
\frac{\delta^{\tilde{\mu}\mu}\delta^{\tilde{\nu}\nu}(D^{\mu\mu}K^{\nu\nu}-K^{\mu\mu}D^{\nu\nu})}{(D^{\mu\mu}D^{\nu\nu}-K^{\mu\mu}K^{\nu\nu})^2}
(P_{j}D-DKP_{j}K+\partial_jD)^{\nu\mu}\\
&=-\frac{\lambda_k-\lambda_l}{(\lambda_k\lambda_l-1)^2}
(\ov{R_{i}}^{kl}\lambda_l-\lambda_k\ov{R_{i}}^{kl}+\partial_i\lambda_k\delta^{kl})(R_{j}^{kl}\lambda_l-\lambda_k R_{j}^{kl}+\partial_j\lambda_k\delta^{kl})\\
&+\frac{\lambda_k-\lambda_l}{(\lambda_k\lambda_l-1)^2}
({R_{i}}^{kl}\lambda_l-\lambda_k{R_{i}}^{kl}+\partial_i\lambda_k\delta^{kl})(\ov{R_{j}}^{kl}\lambda_l-\lambda_k \ov{R_{j}}^{kl}+\partial_j\lambda_k\delta^{kl})\\
&+\frac{\lambda_k+\lambda_l}{(\lambda_k\lambda_l+1)^2}(\ov{Q_{i}}^{kl}\lambda_l+\lambda_k \ov{Q_{i}}^{kl})(Q_{j}^{kl}\lambda_l+\lambda_k Q_{j}^{kl})
-\frac{\lambda_k+\lambda_l}{(\lambda_k\lambda_l+1)^2}(Q_{i}^{kl}\lambda_l+\lambda_k Q_{i}^{kl})(\ov{Q_{j}}^{kl}\lambda_l+\lambda_k \ov{Q_{j}}^{kl})\\
&=-\frac{(\lambda_k-\lambda_l)^3}{(\lambda_k\lambda_l-1)^2}(\ov{R_{i}}^{kl}R_{j}^{kl}-\ov{\ov{R_{i}}^{kl}R_{j}^{kl}})
+\frac{(\lambda_k+\lambda_l)^3}{(\lambda_k\lambda_l+1)^2}(\ov{Q_{i}}^{kl}Q_{j}^{kl}-\ov{\ov{Q_{i}}^{kl}Q_{j}^{kl}})\\
&=-2i\frac{(\lambda_k-\lambda_l)^3}{(\lambda_k\lambda_l-1)^2}\Im[\ov{R_{i}}^{kl}R_{j}^{kl}]+2i\frac{(\lambda_k+\lambda_l)^3}{(\lambda_k\lambda_l+1)^2}\Im[\ov{Q_{i}}^{kl}Q_{j}^{kl}],
\end{split}
\]
which in combination with term $4\partial_i\boldsymbol{d}^\dag\sigma^{-1}K\sigma^{-1}\partial_j\boldsymbol{d}$ gives Eq.~\eqref{eq:multimode_condition}.

Finally, to obtain matrix elements of $\cA_i$ for symmetric logarithmic derivative $\mathscr{L}_i$, we define $\vectorization{W_i}=\mathfrak{M}_{diag}^{-1}\vectorization{\!P_{i}D\!-\!DKP_{i}K\!+\!\partial_iD\!}$, and derive
\[
\begin{split}
\vectorization{\cA_i}&=\mathfrak{M}^{-1}\vectorization{\partial_i\sigma}=((S^{-1})^T\otimes (S^{-1})^\dag)\mathfrak{M}_{diag}^{-1}\vectorization{\!P_{i}D\!-\!DKP_{i}K\!+\!\partial_iD\!}\\
&=((S^{-1})^T\otimes (S^{-1})^\dag)\vectorization{W_i}=\vectorization{(S^{-1})^\dag W_iS^{-1}},
\end{split}
\]
from which we have
\[\label{eq:AvsS}
\cA_i=(S^{-1})^\dag W_iS^{-1}.
\]
Inserting this into the formula for the symmetric logarithmic derivative, Eq.~\eqref{eq:SLD_mixed}, we obtain
\[\label{eq:SLD_mixed3}
\hat{\mathscr{L}}_i(\be)=\dbA^\dag (S^{-1})^\dag W_i S^{-1}\dbA-\frac{1}{2}\tr[DW_i]+2\dbA^\dag\sigma^{-1}\partial_i{\boldsymbol{d}}.
\]

To obtain operator $W_i$ explicitly, using Eq.~\eqref{eq:vecW} we derive
\[
\begin{split}
&W_i^{\nu\mu}=\big(\mathfrak{M}_{diag}^{-1}\big)^{\mu\nu,\alpha\beta}(\!P_{i}D\!-\!DKP_{i}K\!+\!\partial_iD)^{\beta\alpha},\\
&=\frac{\delta^{\mu\alpha}\delta^{\nu\beta}}{D^{\mu\mu}D^{\nu\nu}-K^{\mu\mu}K^{\nu\nu}}
(\!P_{i}D\!-\!DKP_{i}K\!+\!\partial_iD)^{\beta\alpha},\\
&=\frac{\!P_{i}^{\nu\mu}D^{\mu\mu}\!-\!D^{\nu\nu}K^{\nu\nu}P_{i}^{\nu\mu}K^{\mu\mu}\!+\!\partial_iD^{\nu\nu}\delta^{\nu\mu}}{D^{\nu\nu}D^{\mu\mu}-K^{\nu\nu}K^{\mu\mu}}.
\end{split}
\]
We find that $W_i$ is Hermitian, and it has structure
\[\label{eq:structure_of_A}
W_i=\begin{bmatrix}
W_{Xi} & W_{Yi} \\
\overline{W_{Yi}} & \overline{W_{Xi}}
\end{bmatrix},\quad
W_{Xi}^{kl}
=-\frac{\lambda_k-\lambda_l}{\lambda_k\lambda_l-1}R_i^{kl}+\frac{\partial_i\lambda_k}{\lambda_k^2-1}\delta^{kl},\quad
W_{Yi}^{kl}=\frac{\lambda_k+\lambda_l}{\lambda_k\lambda_l+1}Q_i^{kl}.
\]
Finally, we have
\[
\frac{1}{2}\tr[DW_i]=\frac{1}{2}\sum_{\mu,\nu}D^{\nu\mu}W_i^{\mu\nu}=\sum_{k=1}^N\frac{\lambda_k\partial_i\lambda_k}{\lambda_k^2-1}.
\]
which in combination with Eq.~\eqref{eq:SLD_mixed3}, gives Eq.~\eqref{eq:SLD_mixed_williamson}.

\end{widetext}
\section{Problematic behavior at points of purity}\label{app:pops}

It follows from the Heisenberg uncertainty relations that $\lambda_k\geq1$, and $\lambda_k=1$ if and only if the mode is in a pure state. The multi-mode Gaussian state is pure when all symplectic eigenvalues are equal to one. Here we show why terms in Eq.~\eqref{eq:multimode_QFI} that are problematic for pure modes, are defined by Eq.~\eqref{def:problematic_points} for QFIM, and by Eq.~\eqref{def:problematic_points2} for cQFIM respectively.

Let us first take a look at the QFIM. We can ask how to define problematic terms in Eq.~\eqref{eq:multimode_QFI}, which has been derived for $\lambda_k>1$, so that it can also be valid for $\lambda_k=1$.

Rewriting formula for QFIM for pure Gaussian states, Eq.~\eqref{eq:pure_non-elegant}, in terms of the Williamson's decomposition of the covariance matrix yields
\[\label{eq:pure_Williamson}
H^{ij}(\be)=2\sum_{k,l=1}^{N}\Re[\ov{Q_{i}}^{kl}Q_{j}^{kl}]+2\partial_i\boldsymbol{d}^\dag\sigma^{-1}\partial_j\boldsymbol{d}.
\]
Since for pure Gaussian states $\lambda_k(\be)=1$ for all $k$, we can easily see that Eq.~\eqref{eq:multimode_QFI} becomes identical with Eq.~\eqref{eq:pure_Williamson}, when we define problematic terms as
\[
\frac{(\lambda_k\!-\!\lambda_l)^2}{\lambda_k\lambda_l\!-\!1}(\be)=0,\quad
\frac{\partial_i\lambda_k\partial_j\lambda_k}{\lambda_k^2-1}(\be)=0,
\]
for $\be$ such that $\lambda_k(\be)=\lambda_l(\be)=1$.

We assume that defining problematic terms this way is valid also when only some symplectic eigenvalues are equal to 1, but not all of them as in the case of pure Gaussian states. Of course, this could also be precisely derived by carefully studying this intermediate case, for example by using methods introduced in Ref.~\cite{Monras2013a} for a single-parameter estimation: solving the Stein equation for the symmetric logarithmic derivatives, and utilizing the Moore-Penrose pseudoinverse for writing the solution for the QFIM.

Now, according to Ref.~\cite{vsafranek2017discontinuities}, the cQFIM can be calculated from QFIM by performing a special limit:
\[\label{eq:limitHc}
H_c^{ij}(\be)=\lim_{\de\rightarrow 0}H^{ij}(\be+\de\,\boldsymbol{e}_i)=\lim_{\de\rightarrow 0}H^{ij}(\be+\de\,\boldsymbol{e}_j),
\]
where $\boldsymbol{e}_i=(0,\dots,0,1,0\dots,0)$ denotes a unit vector with number $1$ at the $i$'th position.

Let $\be$ be a point such $\lambda_k(\be)=1$ for some $k$, and $\sigma(\be)\in C^{(2)}$. The function $\lambda_k$ must achieve the local minimum at point $\be$, because $\lambda(\be)\geq 1$. The Taylor expansion must be of the form
\[
\lambda_k(\be+\bdeps)=1+\frac{1}{2}\bdeps^T\mathcal{H}_k\bdeps+\cdots,
\]
where $\mathcal{H}_k^{ij}=\partial_i\partial_j\lambda_k$ is the positive semi-definite matrix called Hessian. The Taylor expansion of the derivative at the same point is
\[
\partial_i\lambda_k(\be+\bdeps)=\frac{1}{2}(\boldsymbol{e}_i^T\mathcal{H}_k\bdeps+\bdeps^T\mathcal{H}_k\boldsymbol{e}_i)+\cdots
=\sum_j\mathcal{H}_k^{ij}\bdeps^j+\cdots.
\]
Specifically, for $\bdeps\equiv \de\boldsymbol{e}_i$ we have
\begin{subequations}
\begin{align}
\lambda_k(\be+\de\boldsymbol{e}_i)&=1+\frac{1}{2}\mathcal{H}_k^{ii}\de^2+\cdots,\\
\partial_i\lambda_k(\be+\de\boldsymbol{e}_i)&=\mathcal{H}_k^{ii}\de+\cdots,\\
\partial_j\lambda_k(\be+\de\boldsymbol{e}_i)&=\mathcal{H}_k^{ij}\de+\cdots.
\end{align}
\end{subequations}

According to Eq.~\eqref{eq:limitHc}, and assuming $\be$ is such that $\lambda_k(\be)=\lambda_l(\be)=1$, to obtain the cQFIM $H_c$, the problematic terms in Eq.~\eqref{eq:multimode_QFI} must be defined as
\begin{subequations}
\begin{align}
\frac{(\lambda_k\!-\!\lambda_l)^2}{\lambda_k\lambda_l\!-\!1}(\be)\!&:=\!\lim_{\de\rightarrow 0}\frac{(\lambda_k(\be+\de\,\boldsymbol{e}_i)\!-\!\lambda_l(\be+\de\,\boldsymbol{e}_i))^2}{\lambda_k(\be+\de\,\boldsymbol{e}_i)\lambda_l(\be+\de\,\boldsymbol{e}_i)\!-\!1}\nonumber\\
&=\lim_{\de\rightarrow 0}\frac{\frac{1}{4}(\mathcal{H}_k^{ii}-\mathcal{H}_l^{ii})^2\de^4+\cdots}{\frac{1}{2}(\mathcal{H}_k^{ii}+\mathcal{H}_l^{ii})\de^2+\cdots}=0,\\
\frac{\partial_i\lambda_k\partial_j\lambda_k}{\lambda_k^2-1}(\be)\!&:=\!\lim_{\de\rightarrow 0}\frac{\partial_i\lambda_k(\be+\de\,\boldsymbol{e}_i)\partial_j\lambda_k(\be+\de\,\boldsymbol{e}_i)}{\lambda_k^2(\be+\de\,\boldsymbol{e}_i)-1}\nonumber\\
&=\lim_{\de\rightarrow 0}\frac{\mathcal{H}_k^{ii}\mathcal{H}_k^{ij}\de^2+\cdots}{\mathcal{H}_k^{ii}\de^2+\cdots}=\mathcal{H}_k^{ij}.
\end{align}
\end{subequations}

\section{Estimation of the remainder in the limit formula}\label{app:Remainder}
Here we prove the bound on the remainder of the limit formula, Eq.~\eqref{eq:remainder}. We consider the Williamson decomposition $\sigma=SDS^\dag$. An element of the infinite series, Eq.~\eqref{eq:limit_formula}, can be written as
\[
a_n=\tr\big[A^{-n}\partial_i{A}A^{-n}\partial_j{A}\big]=\tr\big[{D^{-n}B_iD^{-n}B_j}\big],
\]
where $B_i=S^\dag\partial_i{A}(S^\dag)^{-1}K^{-n-1}$. We can derive inequalities
\[\label{eq:Remainder1}
\begin{split}
\norm{a_n}&=\norm{\sum_{k,l}\frac{1}{\lambda_k^n\lambda_l^n}B_i^{kl}B_j^{lk}}
\leq\norm{\sum_{k,l}\frac{1}{\lambda_k^{n}\lambda_l^{n}}\norm{B_i^{kl}}\norm{B_j^{lk}}}\\
&\leq\frac{1}{\lambda_{\mathrm{min}}^{2n}}\norm{\sum_{k,l}\norm{B_i^{kl}}\norm{B_j^{lk}}}\leq\frac{1}{\lambda_{\mathrm{min}}^{2n}}\sqrt{\sum_{k,l}\norm{B_i^{kl}}^2}\sqrt{\sum_{k,l}\norm{B_j^{kl}}^2}\\
&=\frac{1}{\lambda_{\mathrm{min}}^{2n}}\sqrt{\tr[B_i^\dag B_i]}\sqrt{\tr[B_j^\dag B_j]},
\end{split}
\]
where the last inequality is the Cauchy-Schwarz inequality between $B_i^{kl}$ and $B_j^{lk}$ considered as vectors with $2N\times2N$ entries where $N$ is number of modes, $\lambda_{\mathrm{min}}:=\min_{i}\{\lambda_i\}$ is the smallest symplectic eigenvalue. Defining Hermitian matrix $C_i:=S^\dag \partial_i{A}KS$ we have
\[\label{eq:Remainder2}
\begin{split}
\tr[(A\partial_i{A})^2]&=\tr[C_i^\dag D C_i D]=\sum_{k,l}\norm{C_i^{kl}}^2\lambda_k\lambda_l\\
&\geq\lambda_{\mathrm{min}}^2\tr[C_i^\dag C_i]=\lambda_{\mathrm{min}}^2\tr[B_i^\dag B_i].
\end{split}
\]
Combining \eqref{eq:Remainder1} and \eqref{eq:Remainder2} gives
\[
\norm{a_n}\leq\sqrt{\tr\big[(A\partial_i{A})^2\big]}\sqrt{\tr\big[(A\partial_j{A})^2\big]}\lambda_{\mathrm{min}}^{-2n-2}.
\]
For $\lambda_{\mathrm{min}}>1$ we can estimate the remainder,
\[
\begin{split}
\norm{R_M}&\leq \frac{\sqrt{\tr\big[(A\partial_i{A})^2\big]}\sqrt{\tr\big[(A\partial_j{A})^2\big]}}{2}\!\!\sum_{n=M+1}^\infty\!\!\!\!{\lambda_{\mathrm{min}}^{-2n-2}}\\
&=\frac{\sqrt{\tr\big[(A\partial_i{A})^2\big]}\sqrt{\tr\big[(A\partial_j{A})^2\big]}}{2\lambda_{\mathrm{min}}^{2M+2}(\lambda_{\mathrm{min}}^2-1)}.
\end{split}
\]

\section{Expressions for pure states}\label{app:pure_state_condition}
Here we derive Eqs.~\eqref{eq:sld_pure} and~\eqref{eq:pure_state_condition}. We start by deriving
\[
\begin{split}
&\vectorization{\cA_{i\nu}}=\mathfrak{M}_\nu^{-1}\vectorization{\partial_i\sigma}\\
&=(\nu^2\ov{\sigma}\otimes\sigma-K\otimes K)^{-1}\vectorization{\partial_i\sigma}\\
&=\tfrac{1}{\nu^2}(I-(\nu^2\ov{A}\otimes A)^{-1})^{-1}\big(\ov{\sigma}\otimes\sigma\big)^{-1}\vectorization{\partial_i\sigma}\\
&=\tfrac{1}{\nu^2}\sum_{n=0}^\infty(\nu^2\ov{A}\otimes A)^{-n}\big(\ov{\sigma}\otimes\sigma\big)^{-1}\vectorization{\partial_i\sigma}\\
&=\!\tfrac{1}{\nu^2}\bigg(\!\sum_{n=0}^\infty(\nu^2\ov{A}\otimes A)^{-2n}\!+\!(\nu^2\ov{A}\otimes A)^{-2n-1}\!\!\bigg)\big(\ov{\sigma}\otimes\sigma\big)^{-1}\!\!\!\vectorization{\partial_i\sigma}\\
&=\!\tfrac{1}{\nu^2}\!\!\sum_{n=0}^\infty\!\!(I+(\nu^2\ov{A}\otimes A)^{-1})(\nu^{-4}\ov{A}^{-2}\!\otimes\! A^{-2})^{n}\big(\ov{\sigma}\otimes\sigma\big)^{-1}\!\!\!\vectorization{\partial_i\sigma}\\
&=\tfrac{1}{\nu^2}(I+(\nu^2\ov{A}\otimes A)^{-1})\sum_{n=0}^\infty(\nu^{-4})^{n}\big(\ov{\sigma}\otimes\sigma\big)^{-1}\vectorization{\partial_i\sigma}\\
&=\tfrac{\nu^2}{\nu^4-1}(I+(\nu^2\ov{A}\otimes A)^{-1})\big(\ov{\sigma}\otimes\sigma\big)^{-1}\vectorization{\partial_i\sigma}\\
&=\tfrac{\nu^2}{\nu^4-1}(I+\tfrac{1}{\nu^2}(\ov{A}\otimes A)^{-1})\big(\ov{A}\otimes A\big)^{-1}\big(K\otimes K\big)\vectorization{\partial_i\sigma}\\
&=\tfrac{\nu^2}{\nu^4-1}(\ov{A}^{-1}\otimes A^{-1}+\tfrac{1}{\nu^2})\big(K\otimes K\big)\vectorization{\partial_i\sigma}\\
&=\tfrac{\nu^2}{\nu^4-1}(\ov{A}^{-1}\otimes A^{-1}+\tfrac{1}{\nu^2})\vectorization{\partial_iA K}\\
&=\tfrac{\nu^2}{\nu^4-1}(\vectorization{A^{-1}\partial_iA K (A^{-1})^\dag}+\tfrac{1}{\nu^2}\vectorization{\partial_iA K})\\
&=\tfrac{\nu^2}{\nu^4-1}(-\vectorization{\partial_iA A^{-1} K (A^{-1})^\dag}+\tfrac{1}{\nu^2}\vectorization{\partial_iA K})\\
&=\tfrac{\nu^2}{\nu^4-1}(-\vectorization{\partial_iA A^{-2}K}+\tfrac{1}{\nu^2}\vectorization{\partial_iA K})\\
&=\tfrac{\nu^2}{\nu^4-1}(-1+\tfrac{1}{\nu^2})\vectorization{\partial_iA K}\\
&=-\tfrac{1}{1+\nu^2}\vectorization{\partial_iA K},
\end{split}
\]
where we have used $A^2=1$ and $A^{-1}\partial_iA =-\partial_iA A^{-1}$, which holds for pure states.

Taking the limit $\nu\rightarrow 1$ we obtain expression for pure states
\[
\cA_i=-\tfrac{1}{2}\partial_iA K,
\]
which after inserting into Eq.~\eqref{eq:SLD_mixed} gives
\[
\hat{\mathscr{L}}_i(\be)=-\tfrac{1}{2}\dbA^\dag \partial_iA K\dbA+\tfrac{1}{4}\tr[A \partial_iA]+2\dbA^\dag\sigma^{-1}\partial_i{\boldsymbol{d}}.
\]
The second term is zero, because $\tr[A \partial_iA]=\frac{1}{2}\partial_i\tr[A^2]=\frac{1}{2}\partial_i\tr[I]=0$.

Alternatively, we can use
\[
\partial_iA =\partial_iAA^{-2}=-A^{-1}\partial_iAA^{-1}=-\sigma^{-1}\partial_i\sigma \sigma^{-1} K
\]
to derive
\[
\hat{\mathscr{L}}_i(\be)=\tfrac{1}{2}\dbA^\dag\!\sigma^{-1}\! \partial_i\sigma \sigma^{-1}\!\dbA+2\dbA^\dag\!\sigma^{-1}\partial_i{\boldsymbol{d}},
\]
which proves Eq.~\eqref{eq:sld_pure}.

Further,
\[
\begin{split}
&\vectorization{\partial_i\sigma}^\dag\mathfrak{M}_{\nu}^{-1}(\ov{\sigma}\!\otimes\! K\!-\!K\!\otimes\!\sigma)\mathfrak{M}_{\nu}^{-1}\vectorization{\partial_j\sigma}\\
&=\tfrac{1}{(1+\nu^2)^2}\vectorization{\partial_iA K}^\dag(\ov{\sigma}\!\otimes\! K\!-\!K\!\otimes\!\sigma)\vectorization{\partial_jA K}\\
&=\tfrac{1}{(1+\nu^2)^2}\vectorization{\partial_iA K}^\dag(\vectorization{K\partial_jA A}-\vectorization{\sigma\partial_jA})\\
&=\tfrac{1}{(1+\nu^2)^2}\tr[(\partial_iA K)^\dag({K\partial_jA A}-{\sigma\partial_jA})]\\
&=\tfrac{1}{(1+\nu^2)^2}\tr[\partial_iA K({K\partial_jA A}-{\sigma\partial_jA})]\\
&=\tfrac{1}{(1+\nu^2)^2}\tr[A[\partial_iA,\partial_jA]]\\
&=\tfrac{1}{(1+\nu^2)^2}\tr[K\sigma[K\partial_i\sigma,K\partial_j\sigma]],
\end{split}
\]
which in combination with the regularization procedure, Eq.~\eqref{eq:condition_QCRregularization_procedure2}, proves Eq.~\eqref{eq:pure_state_condition}.

\bibliographystyle{apsrev4-1}
\bibliography{multibib}

\end{document}